\documentclass[reprint, superscriptaddress, amsmath, amssymb, aps, nofootinbib]{revtex4-2}
\usepackage[utf8]{inputenc}
\usepackage{amsmath}
\usepackage{amsthm}
\usepackage[normalem]{ulem}
\usepackage{graphicx}
\usepackage{soul}
\usepackage{ulem}
\usepackage{xcolor}
\usepackage[T1]{fontenc}
\usepackage{mathtools}
\graphicspath{{pictures/}}

\begin{document}
\title{Dynamical probe of the pseudo Jahn-Teller effect\\
in one-dimensional confined fermions}
\date{\today}
\author{A. Becker}
\email{andre.becker@uni-hamburg.de}
\affiliation{Center for Optical Quantum Technologies, Department of Physics, University of Hamburg, 
Luruper Chaussee 149, 22761 Hamburg Germany}
\affiliation{The Hamburg Centre for Ultrafast Imaging,
University of Hamburg, Luruper Chaussee 149, 22761 Hamburg, Germany}
\author{G. M. Koutentakis}
\email{georgios.koutentakis@ist.ac.at}
\affiliation{Institute of Science and Technology Austria (ISTA), am Campus 1,\\ 3400 Klosterneuburg, Austria}
\author{P. Schmelcher}
\email{peter.schmelcher@uni-hamburg.de}
\affiliation{Center for Optical Quantum Technologies, Department of Physics, University of Hamburg, 
Luruper Chaussee 149, 22761 Hamburg Germany}
\affiliation{The Hamburg Centre for Ultrafast Imaging,
University of Hamburg, Luruper Chaussee 149, 22761 Hamburg, Germany}

\begin{abstract}
We investigate the real-time dynamics of a quenched quantum impurity immersed in a one-dimensional ultracold Fermi gas, focusing on the breakdown of the adiabatic Born-Oppenheimer approximation due to non-adiabatic effects. Despite a sizable impurity-bath mass imbalance, increasing interactions induce strong non-adiabatic couplings, disrupting adiabatic motion and enabling population transfer between the adiabatic potential energy curves. These transitions are governed by conical intersections arising from the pseudo Jahn-Teller effect, dynamically shaping the impurity’s motion through the bath. Using \textit{ab initio} simulations via the Multi-Layer Multi-Configuration Time-Dependent Hartree method and a multi-channel Born-Oppenheimer framework, we track the impurity’s evolution and directly prove the dynamical manifestation of the pseudo Jahn-Teller effect. We analyze two key scenarios: (i) a small initial shift, where a single avoided crossing drives transitions, and (ii) a large shift, where multiple avoided crossings lead to enhanced non-adiabaticity, self-trapping, and energy redistribution. Our findings establish ultracold fermionic few-body systems as tunable platforms for studying non-adiabatic quantum dynamics, opening new avenues for controlled impurity transport in strongly correlated environments.
\end{abstract}
\maketitle
\section{Introduction}

Non-adiabatic phenomena, emerging when the Born-Oppenheimer approximation ceases to be valid, constitute a crucial area of research in quantum physics. These effects manifest across diverse domains, including molecular dynamics, condensed matter physics, and quantum materials \cite{domcke2012conical, tully1998mixed, zhu2015charge, cappelluti1997nonadiabatic, hasan2010topological, dalibard2011artificial}. The breakdown of the Born-Oppenheimer approximation occurs when the fast and slow degrees of freedom that typically enable an adiabatic separation become strongly correlated. This correlation results in intricate dynamical behaviors that govern fundamental processes such as energy redistribution, symmetry breaking, and quantum coherence, particularly in photochemical reactions \cite{Koeppel1983, Koizumi1999, Koizumi2000, Domcke2004, woener_2009, Demekhin2013, Lasorne2014, Galbraith2017, CrespoOtero2018, mukherjee_beyond_2019, Bersuker2021}. The (pseudo-)Jahn-Teller effect provides a well-established paradigm for non-adiabatic coupling in systems exhibiting (or lacking) energetic degeneracy \cite{JahnTeller1937, Englman1972, Bersuker2006, Köppel2009, Bersuker2021}. This phenomenon involves vibronic distortions driven by interactions between (nearly) degenerate electronic states and vibrational modes, leading to symmetry breaking and significantly altering the system's dynamical properties \cite{BersukerPolinger1989, Bersuker2013}.
\begin{figure*}
    \centering
    \includegraphics[width=1\linewidth]{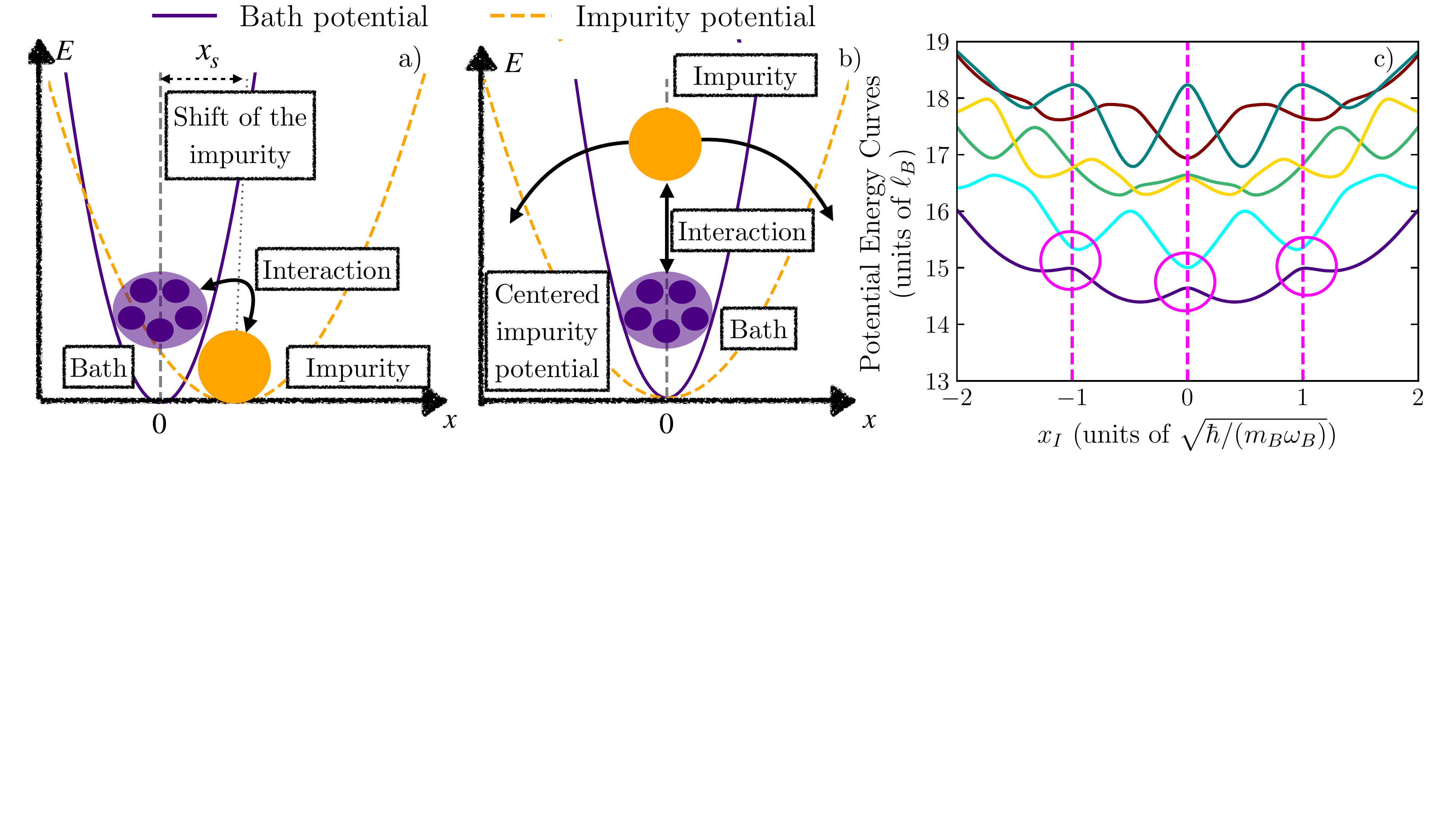}
    \caption{(a) Schematic illustration of the initial prequench configuration of our many-body system featuring a shifted impurity trap relative to its fermionic bath. (b) To initiate the dynamics we perform a quench of the trapping potential to align the bath and impurity confinement. (c) PEC \eqref{eqn:potential_energy_surfaces} from the MCBO approach for $g=5.0~\hbar\omega_B\ell_B$ with the trapping parameters $m_I=4m_B$ and $\omega_I=0.5\omega_B$~\cite{Becker2024}. The magenta-colored circles indicate the positions of the avoided crossings associated to the PJTE induced synthetic conical intersections at $g \to \infty$, and the dashed line indicates their positions.}
    \label{fig:setup_and_potential_energy_surfaces}
\end{figure*}

Historically, non-adiabaticity was first studied in molecular physics before extending into condensed matter systems \cite{BornOppenheimer1927, born1954dynamical}. However, the experimental challenges associated with controlling interactions and observing real-time dynamics have driven the search for alternative platforms. Ultracold atomic systems have emerged as a promising area for investigating non-adiabatic effects \cite{Tiemann2020, Kendrick2021, Pant2021}. Atoms cooled and confined in optical trapping potentials enable precise experimental simulations of molecular dynamics, providing new opportunities to study fundamental quantum phenomena. In particular, fermionic few-body systems exhibit pronounced non-adiabatic effects due to strong correlations and low-dimensional confinement. These setups allow for an in-depth exploration of interactions governing quantum correlations, phase transitions, and emergent collective phenomena \cite{Zuern2012, Wenz_2013, Murmann2015, Bayha2020, Holten2021}.

In ultracold fermionic few-body systems, non-adiabatic couplings manifest as avoided crossings in the potential energy surfaces \cite{Becker2024}. The ability to tune the system parameters using external fields, such as magnetic fields to leverage Fano-Feshbach and confinement-induced resonances, as well as species-selective trapping geometries, provides an unprecedented opportunity to study vibronic couplings and pseudo Jahn-Teller effect (PJTE)-like distortions in a context distinct from molecular physics \cite{Voigt2009, Tiecke2010, Naik_2011, Cetina2015}. These tunable systems offer a bridge between fundamental studies of non-adiabatic physics and experimentally realizable dynamic behaviors in engineered quantum systems.

In our comprehensive ground-state analysis \cite{Becker2024}, we thoroughly investigated the spatial structure of the potential energy curves (PEC) of a one-dimensional harmonically confined fermionic system composed of an impurity and a Fermi gas. By introducing the interaction strength as a synthetic dimension, we discovered the emergence of multiple conical intersections in the infinite interaction limit, which were attributed to the PJTE. For finite repulsions these conical intersections introduce avoided crossings among the PEC that lead to pronounced non-adiabatic effects in the ground state of the system.

In this work, we delve deeper into the significance of these PJTE induced avoided crossings, particularly regarding the system's dynamics. To achieve that we initialize the system in its ground state with a shifted impurity confinement and perform a quantum quench to revert this shift. To track the subsequent dynamics, we evolve the system using our numerically exact, {\it ab initio} tool for exploring dynamic quantum systems, the Multi-Layer Multi-Configuration Time-Dependent Hartree method for bosons and fermions (ML-X) \cite{CaoLushuai2013, CaoBolsinger2017}. A comparative analysis with our multi-channel Born-Oppenheimer (MCBO) methodology reveals the influence of the PJTE on the time evolution of the system. Through our detailed investigation of the system’s dynamics, we provide direct evidence that the PJTE governs the interplay between non-adiabatic transitions and strong correlations. By identifying the dominant avoided crossings responsible for population transfer between different PECs, we conclusively demonstrate how the PJTE dictates the dynamical evolution of the system. In this study, we analyze the dynamics in two decisive scenarios: first, a small shift of the center of the impurity species, where we primarily focus on the contribution of a single avoided crossing between the two lowest PECs at the center of the trap (see Fig.~\ref{fig:setup_and_potential_energy_surfaces}(c)). 
Second, in the regime of a larger shift, where additional avoided crossings in the outer region, together with the central avoided crossing, strongly enhance non-adiabatic effects, leading to the self-trapping of the impurity and the emergence of collective excitations of the bath species.

Our work is structured as follows. Section \ref{sct:Setup} introduces the underlying PJTE framework and presents a one-dimensional Hamiltonian describing a few-body fermionic bath interacting with an impurity via s-wave interactions. Section \ref{sec:Methodology and Computation Approach} details the \textit{ab initio} ML-X and MCBO approaches employed to solve the time-dependent Schrödinger equation. 
In Section \ref{sec:Time-dependent behavior of the one-body density}, we focus on the impact of different initial shifts of the impurity species trapping potential on the system's time evolution, emphasizing how these shifts influence the manifestation of non-adiabatic effects. Section \ref{sec:Stronger trapping confinement of the impurity} examines how stronger trapping confinement modifies the observed dynamics, providing further evidence for the PJTE in the studied system.
We conclude with a summary of our findings and an outlook on future research directions in Section \ref{sec:Summary and Outlook}.
Additional technical details are provided in the appendices. Appendix \ref{app:MLX} elaborates on the ML-X formalism, which enables precise dynamical studies \cite{KrönkeCao2013, CaoLushuai2013, CaoBolsinger2017}. Finally, Appendix \ref{app:Derivations in Second Quantization Formalism} contains derivations of PEC occupation observables using the second quantization formalism.

Our findings provide crucial insights into the interplay between strong correlations, non-adiabaticity, and vibronic interactions in ultracold fermionic systems. By delineating the regimes where synthetic conical intersections significantly influence the dynamics, our results offer a practical guide for future few-body experiments, paving the way for controlled studies of emergent quantum behaviors in engineered quantum matter.
\section{Setup}
\label{sct:Setup}

We investigate a two-species setup of mass-imbalanced and spin-polarized fermions, both confined in individual one-dimensional harmonic trapping potentials. The system consists of a lighter majority species, denoted as the bath ($B$), which interacts with a single heavier impurity ($I$), i.e., $m_I \gg m_B$, via s-wave repulsion.

Building upon our previous ground-state analysis~\cite{Becker2024}, we consider the quench scenario schematically depicted in Fig.~\ref{fig:setup_and_potential_energy_surfaces}. 
As shown in Fig.~\ref{fig:setup_and_potential_energy_surfaces}(a), the system is first prepared by shifting the harmonic trapping potential of the impurity by an initial displacement $x_s$. The ground state of this displaced system is computed before the trapping potential is abruptly quenched to the center ($x_s = 0$) at $t=0$, as illustrated in Fig.~\ref{fig:setup_and_potential_energy_surfaces}(b). This sudden quench initiates a dipole oscillation of the impurity, allowing us to study the interplay between its motion and the bath response.

We analyze two distinct displacement scenarios: a small shift ($x_s = 0.5\ell_B$) and a large shift ($x_s = 1.5\ell_B$). As demonstrated in Fig.~\ref{fig:setup_and_potential_energy_surfaces}(c), these shifts determine how the impurity traverses the avoided crossings (associated with conical intersections emerging in the limit of infinite repulsion) during its dynamics. The large shift case, in particular, results in the impurity crossing multiple avoided crossings, amplifying non-adiabatic effects and self-trapping phenomena. The avoided crossings are manifestations of the PJTE and will be substantial for the dynamical proof of the PJTE. 

The system is described by the Hamiltonian:
\[ \hat{H} = \hat{H}_B+\hat{H}_I+\hat{H}_{BI}\]
where the individual terms are given by
\begin{equation}
    \begin{split}
    \hat{H}_{B}&=\sum_{j=1}^{N_B}\left[ -\frac{\hbar^2}{2m_B}\left(\frac{\partial}{\partial x_j^B}\right)^2+ \frac{1}{2}m_B \omega_B^2 (x_j^{B})^2 \right],\\
    \hat{H}_{I}&=\sum_{j=1}^{N_I}\left[ -\frac{\hbar^2}{2m_I}\left(\frac{\partial}{\partial x_j^I}\right)^2+ \frac{1}{2}m_I\omega_I^2(x_j^{I}-x_s)^2\right],\\
    \hat{H}_{BI}&=\sum_{k=1}^{N_B}\sum_{j=1}^{N_I}g\delta(x_k^B-x_j^I).
    \end{split}
\end{equation}

Here, $m_B$ and $m_I$ denote the masses of the bath and impurity species, respectively, while $\omega_B$ and $\omega_I$ represent their corresponding trapping frequencies. Our primary focus is on a heavy impurity with mass $m_I = 4 m_B$ and a weakly confined impurity potential $\omega_I = 0.5\omega_B$. This setup is experimentally feasible in state-of-the-art systems, such as $^{6}$Li-$^{23}$Na mixtures \cite{Taglieber2008}. $N_B$ represents the number of bath atoms and $N_I$ the number of impurity atoms.
The characteristic length scale of the bath species is given by the harmonic oscillator length 
\begin{equation}
    \ell_B=\sqrt{\frac{\hbar}{m_B\omega_B}},
\end{equation}
which describes the spatial extent of a non-interacting single-particle ground state. Similarly, the impurity’s spatial extent is characterized by
\begin{equation}
    \ell_I=\sqrt{\frac{m_B\omega_B}{m_I\omega_I}}\ell_B.
\end{equation}
Since the bath consists of multiple fermions, its density distribution extends beyond $\ell_B$ due to the occupancy of higher-energy states. Specifically, the size of the bath is approximately given by
\begin{equation}
    \sigma_B\approx\ell_B\sqrt{N_B+\frac{1}{2}}.
\end{equation}
For investigating non-adiabatic behavior, it is particularly insightful to consider cases where the impurity delocalization matches this scale, i.e.,
\begin{equation}
    \ell_I \approx \sqrt{N_B}\ell_B,
\end{equation}
which grows with the energy of the bath's highest occupied state.
When the length scales $\sigma_B$ and $\ell_I$ are almost equal, the impurity can better probe the bath dynamics, enhancing the system's sensitivity to observe non-adiabatic effects. In this work, we fix the bath particle number to $N_B = 5$ and consider a single impurity ($N_I = 1$).
Due to the impurity's large mass, a Born-Oppenheimer-like approach is justified. In Section~\ref{sec:Stronger trapping confinement of the impurity}, we provide an analysis for the tightly trapped case ($\omega_I \gg \omega_B$).

The impurity interacts with the bath via a contact interaction of strength $g$, modeled by a Dirac delta potential. Importantly, intra-species interactions are excluded due to the Pauli exclusion principle, preventing identical fermions from occupying the same quantum state. In ultracold gas experiments, the interaction strength $g$ is tunable via confinement-induced resonances and Fano-Feshbach (in our case interspecies) resonances, allowing precise control over non-adiabatic effects \cite{ChinChengGrimm2010}.

\section{Methodology and computational approach}
\label{sec:Methodology and Computation Approach}
In the following, we introduce the numerical methods for our analysis:
We begin with our fully correlated ML-X method. Next, we address our MCBO approach, which is motivated by the significant mass imbalance between bath and impurity species \(m_I/m_B > 1\). Both methods are numerically exact \textit{ab-initio} approaches suitable for the solution of multi-component fermionic systems. 
\subsection{The ML-X method}
The Multi-Layer Multi-Configuration Time-Dependent Hartree method for mixtures (ML-X) is a variational, \textit{ab initio}, and numerically exact approach to simulate the non-equilibrium quantum dynamics of bosonic and fermionic particles, as well as their mixtures \cite{KrönkeCao2013, CaoLushuai2013, CaoBolsinger2017}. It employs a multi-layered ansatz that variationally optimizes the quantum basis across multiple structural levels of the total many-body wavefunction. This expansion adapts dynamically to inter-particle correlations at the single-particle, single-species, and multi-species levels, enabling efficient and accurate simulations.

The total wavefunction, 
\begin{equation}
|\Psi(t)\rangle = \sum_{k=1}^D \sqrt{\lambda_k(t)} |\tilde{\Psi}_k^B(t)\rangle |\tilde{\Psi}_k^I(t)\rangle,
\end{equation}
is represented using a Schmidt decomposition of rank \( D \). This decomposition expands the many-body wavefunction in terms of single-species functions \( |\tilde{\Psi}_k^{\sigma}(t)\rangle \), where \( \sigma \) denotes the species. The Schmidt weights, \( \lambda_k(t) \), quantify the entanglement between different species, providing a compact representation of the system dynamics.
Each single species function \( |\tilde{\Psi}_k^{\sigma}(t)\rangle \) is expanded in terms of Fock states constituted from \( d^{\sigma} \) time-dependent single-particle functions (SPFs), \( |\phi_i^{\sigma}(t)\rangle \), which dynamically adapt to the evolving system. These SPFs are further expressed in a discrete variable representation (DVR) basis, enabling an efficient and precise representation of the spatial degrees of freedom.
A detailed derivation of this multi-layer representation, including the role of the hierarchical structure of the ansatz, can be found in Appendix~\ref{app:MLX}.

Ground states are determined via imaginary-time propagation. The Hilbert space truncation is characterized by the configuration $C = (D; d^B; d^I)$, chosen to properly account for interspecies entanglement and bath state occupation.

\subsection{The multi-channel Born-Oppenheimer approach}
\label{mcBO_main_text}

Motivated by the assumption of a heavy, less mobile impurity, the comparison of our results with a Born-Oppenheimer-like approach is justified. A general variational formulation of this approach can be established in terms of the MCBO ansatz,
\begin{equation}
\begin{split}
\Psi(x^B_1,\dots,x^B_{N_B}, x_I ; t) &= \\ \sum_{j = 1}^M \Psi_{j,I}(x_I ; t) 
&\underbrace{\Psi_{j,B}(x^B_1,\dots,x^B_{N_B};x_I)}_{\equiv\langle x^B_1,\dots,x^B_{N_B} | \Psi_{j,B}(x_I) \rangle}.
\end{split}
\label{eqn:multi-channel_BornOppenheimer}
\end{equation}
Here, we have introduced an orthonormal basis for the bath species, \(| \Psi_{j,B} (x_I) \rangle\), with \(j=1,2,\dots\), exhibiting a parametric dependence on \(x_I\). 
The impurity-species wavefunctions \(\Psi_{j,I}(x_I ; t)\) with the normalization condition \(\sum_{j = 1}^M \int \mathrm{d}x_I |\Psi_{j,I}(x_I ; t)|^{2} = 1\) correspond to the expansion coefficients in the many-body basis of the coupled system.

Considering a time-dependent variational principle, such as the Dirac-Frenkel one, $\langle \delta \Psi| \hat{H} - i \hbar \frac{\partial}{\partial t} | \Psi \rangle = 0$ , and using the ansatz of Eq.~\eqref{eqn:multi-channel_BornOppenheimer}, we can derive the time-dependent equation of motion
\begin{equation}
    \begin{split}
i\hbar \frac{\mathrm d}{\mathrm d t}\Psi_{k,I}(x_I,t) &= - \frac{\hbar^2}{2 m_I} \sum_{j,l = 1}^M \left( \delta_{kj} \frac{\mathrm{d}}{\mathrm{d}x_{I}} -i A_{kj}(x_I) \right) \\
& \hspace{0.2cm} \times \left( \delta_{jl} \frac{\mathrm{d}}{\mathrm{d}x_{I}} -i A_{jl}(x_I) \right) \Psi_{l,I}(x_I) \\
& + \sum_{l = 1}^M \bigg( \delta_{k l} \varepsilon_k (x_I) + \delta_{kl} \frac{1}{2} m_B \omega^2_I x_I^2 \\
&  \hspace{0.2cm} + V^{\rm ren}_{kl}(x_I) \bigg) \Psi_{l,I}(x_I, t).
\label{eqn:schroedinger-time-dendent}
    \end{split}
\end{equation} 
where the states \(|\Psi_{k,B}(x_I)\rangle\) can, in principle, be any complete wavefunction basis. A natural and practical choice is the eigenstates of \(\hat{H}_B + \hat{H}_{BI}\) for fixed \(x_I\), yielding the diagonal matrix elements  
\begin{equation}
    \langle \Psi_{k,B}(x_I) | \hat{H}_B + \hat{H}_{BI} | \Psi_{l,B}(x_I) \rangle = \delta_{kl} \varepsilon_k(x_I),
    \label{eqn:potential_energy_surfaces}
\end{equation}  
with the PECs given by \(\varepsilon_k(x_I)\), which represent the energy of the bath for a fixed position of the impurity \(x_I\). 
These curves define the adiabatic potential energy landscape through which the impurity moves.
Furthermore, \(A_{kj}(x_I) = i \langle \Psi_{k,B}(x_I) | \frac{\partial \Psi_{j,B}}{\partial x_I} (x_I) \rangle\) is the non-adiabatic derivative coupling, it measures how the bath's eigenstates change with the impurity position arising from the impurity-bath coupling. 
The potential renormalization is  
\begin{equation}
    V_{kl}^{\rm ren}(x_I) = \frac{\hbar^2}{2 m_I} \left\langle \tfrac{\mathrm{d} \Psi_{k,B}}{\mathrm{d}x_I}(x_I) \middle| 1 - \hat{\mathcal{P}}_M \middle| \tfrac{\mathrm{d} \Psi_{l,B}}{\mathrm{d}x_I}(x_I) \right\rangle,
    \label{potential_renormalization}
\end{equation}  
with the projector \(\hat{\mathcal{P}}_M\) onto the subspace spanned by \(|\Psi_{k,B}(x_I)\rangle\) given by  
\begin{equation}
    \hat{\mathcal{P}}_M = \sum_{j=1}^M | \Psi_{j,B}(x_I) \rangle \langle \Psi_{j,B}(x_I) |. 
\end{equation} 
Physically, the potential renormalization represents how the motion of the impurity causes changes in the bath's kinetic energy, as well as subtle adjustments in the bath-impurity interaction that are not captured by the adiabatic approximation alone.
For details on the derivation of the above equations in the static case see Ref.~\cite{Becker2024}.

Before proceeding let us comment on the reduction of the above to the adiabatic Born-Oppenheimer approximation widely employed in molecular physics~\cite{BornOppenheimer1927}.
If we restrict the ansatz of Eq.~\eqref{eqn:multi-channel_BornOppenheimer} to $M = 1$ term, the variational equations of motion reduce to the adiabatic Born-Oppenheimer approximation incorporating the Born-Huang correction arising from $V^{\rm ren}_{11}(x_I)$, which is therefore characterised as variational adiabatic Born-Oppenheimer (VABO) approximation. The usual adiabatic Born-Oppenheimer approach consists of dropping this additional term by considering $V_{11}^{\rm ren}(x_I) = 0$ and will be denoted as non-variational adiabatic Born-Oppenheimer (NVABO) approximation.
\begin{figure*}
    \centering
    \includegraphics[width=0.9\linewidth]{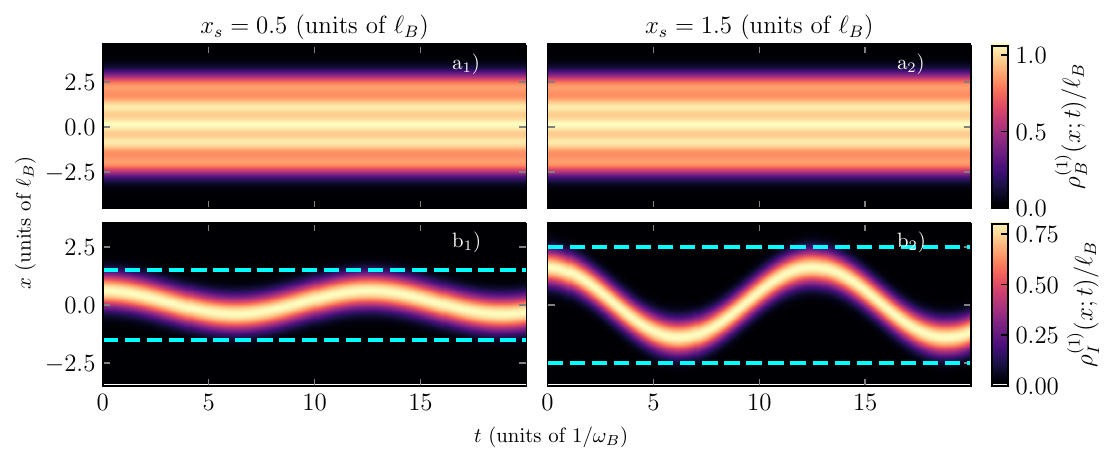}
    \caption{Spatiotemporal evolution of the one-body density of the bath $\rho_B^{(1)}(x;t)$ and impurity $\rho_I^{(1)}(x;t)$ species for a non-interacting system. The impurity mass $m_I=4m_B$ and trapping frequency $\omega_I=0.5\omega_B$ are taken into account. The different columns refer to different shifts (see labels). For the impurity species, we include cyan dashed lines at $x = \pm(x_s+\sqrt{2}\ell_I)$ as a visual reference to the oscillation amplitude of the impurity.
    }
\label{fig:one_body_density_g00}
\end{figure*}

\textcolor{black}{The full MCBO ansatz captures the complete interplay between the impurity and bath species. In contrast, the VABO and NVABO approximations describe the impurity's dynamics within a single-channel potential. By including all non-adiabatic couplings between bath states and a renormalized potential that accounts for transitions driven by impurity motion, MCBO provides a framework for studying non-adiabatic effects in the system. The approximations, VABO and NVABO, are valid in regimes where a single bath state dominates, and non-adiabatic effects are minimal.
\textcolor{black}{}
A key distinction between the two approximations lies in their treatment of the Born-Huang correction. This term which appears in the VABO approximation, captures the variation in the kinetic energy of the bath as a function of the impurity position, reflecting how the bath's response changes with the impurity's motion. This term can influence the dynamics, particularly in cases where the interaction between the impurity and the environment is non-negligible and varies with the impurity’s position. By incorporating or neglecting this correction, we are able to examine different levels of approximation for describing these complex interactions, and thus, VABO and NVABO offer valuable insights into the accuracy and behavior of models that describe such systems.}

\section{Dynamics in the case of weak confinement $\ell_I \approx \sqrt{N_B} \ell_B$}
\label{sec:Time-dependent behavior of the one-body density}

\begin{figure*}
    \centering
    \includegraphics[width=0.9\linewidth]{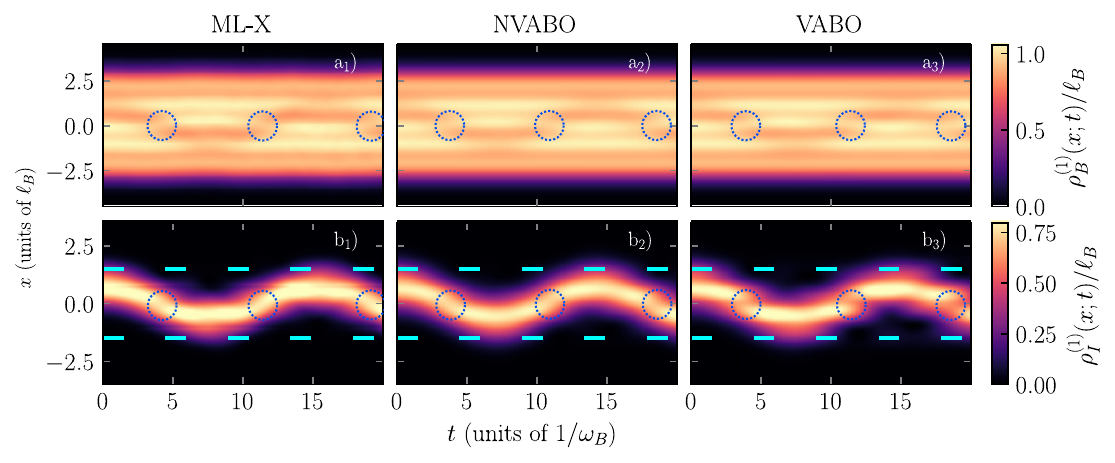}
    \caption{Spatiotemporal dependence of the one-body density of the bath $\rho_B^{(1)}(x;t)$ and impurity $\rho_I^{(1)}(x;t)$ species for an interacting system with $g=2.5\hbar\ell_B\omega_B$. The impurity mass $m_I=4m_B$ and trapping frequency $\omega_I=0.5\omega_B$ as well as the smaller shift $x_s=0.5\ell_B$ are considered. The different columns correspond to the distinct methods (see labels). For the impurity species, we include cyan dashed lines at $x = \pm(x_s+\sqrt{2}\ell_I)$ as a visual reference to the oscillation amplitude in the non-interacting case. \textcolor{black}{The dotted blue circles indicate the spatiotemporal region for which the impurity crosses $x_I = 0$, corresponding to the location of an avoided crossing.}}
    \label{fig:one_body_density_g25_x05}
\end{figure*}
As discussed in Sec.~\ref{sct:Setup}, the relation between the length scales of the bath and impurity species has profound impact on the physics of the system. In this section we focus on the dynamics in the case of a weak trapping potential for the impurity, where the non-adiabatic effects are expected to be pronounced.

In particular, as shown in Fig.~\ref{fig:setup_and_potential_energy_surfaces}(c) and supported by our comprehensive ground-state analysis \cite{Becker2024}, weak trapping potentials give rise to multiple synthetic conical intersections in the $g \to \infty$ limit. The corresponding avoided crossings at finite $g$ are significantly more pronounced compared to systems with higher trapping frequencies, where the impact of conical intersections is noticeable only as the \(g\to\infty\) limit is approached for \(g \gg \hbar \omega_B \ell_B \). The influence of the conical intersections serves as a hallmark of the PJTE, whose profound impact on the system's dynamics will be analyzed in detail in the following.

The time-dependent one-body density is a pivotal quantity that connects theoretical predictions with experimental observations. It offers a simplified, yet comprehensive representation of the dynamical behavior of our system for both the bath (\(\rho_B^{(1)}(x;t)\)) and impurity (\(\rho_I^{(1)}(x;t)\)) species, where \(\rho_{\sigma}^{(1)}(x;t)=\langle\Psi(t)|\hat{\Psi}_{\sigma}^{\dagger}(x)\hat{\Psi}_{\sigma}(x)|\Psi(t)\rangle\) with \(|\Psi(t)\rangle\)  being the many-body wavefunction at time \(t\) and \(\hat{\Psi}_{\sigma}^{\dagger}(x)\) and \(\hat{\Psi}(x)\) are the field operators creating and annihilating a $\sigma$-species, $\sigma \in \{ B, I\}$, particle at position \(x\) respectively.

In the following we would compare two scenarios: a smaller shift given by $x_s=0.5\ell_B$ and a larger shift $x_s=1.5\ell_B$. 
To get an initial overview for the non-interacting case $g=0$, we present the spatio-temporal density evolution in Fig.~\ref{fig:one_body_density_g00}. 
Notice that in this case all of the employed methods ML-X, VABO, NVABO and MCBO provide identical results since the impurity and bath degrees of freedom are separable.
In both cases of $x_s = 0.5 \ell_B$,~Fig.~\ref{fig:one_body_density_g00}(b$_1$), and $x_s = 1.5 \ell_B$,~Fig.~\ref{fig:one_body_density_g00}(b$_2$), we observe that the density of the impurity exhibits a dipole mode with frequency $\omega_I$. However, for $x_s = 1.5 \ell_B$ the amplitude of the oscillatory pattern is increased since more energy is imparted to the system.
The bath species remains unperturbed, see Fig.~\ref{fig:one_body_density_g00}(b$_1$) and~\ref{fig:one_body_density_g00}(b$_2$), since there is no bath impurity interaction, $g = 0$. Its density $\rho_B^{(1)}(x;t)$ exhibits five peaks owing to the Pauli principle resulting in the occupation of the five lowest harmonic oscillator states due to the considered particle number of the bath species ($N_B = 5$).

Considering the interacting case we expect substantial difference between the above mentioned cases of $x_s$ values. For the smaller shift (\(x_s = 0.5 \ell_B\)), the dynamics involves the impurity crossing through a single avoided crossing of the lowest PEC, at $x = 0$, see Fig.~\ref{fig:setup_and_potential_energy_surfaces}(c) resulting in limited non-adiabatic contributions. In contrast, the larger shift (\(x_s = 1.5 \ell_B\)) involves multiple avoided crossings, significantly enhancing the role of non-adiabatic effects driven by the PJTE (see Fig.~\ref{fig:setup_and_potential_energy_surfaces}(c)). These differences necessitate a two-part investigation to capture the unique physics of each scenario.

\subsection{Role of interactions for the smaller shift}

We first analyze the effects of the interactions for the smaller shift. We begin by examining the one-body densities of the bath and impurity species and then proceed with a detailed analysis of the involved PEC and their occupations.

\subsubsection{Density analysis}
\label{density_analysis_small}

\begin{figure*}
    \centering
    \includegraphics[width=0.9\linewidth]{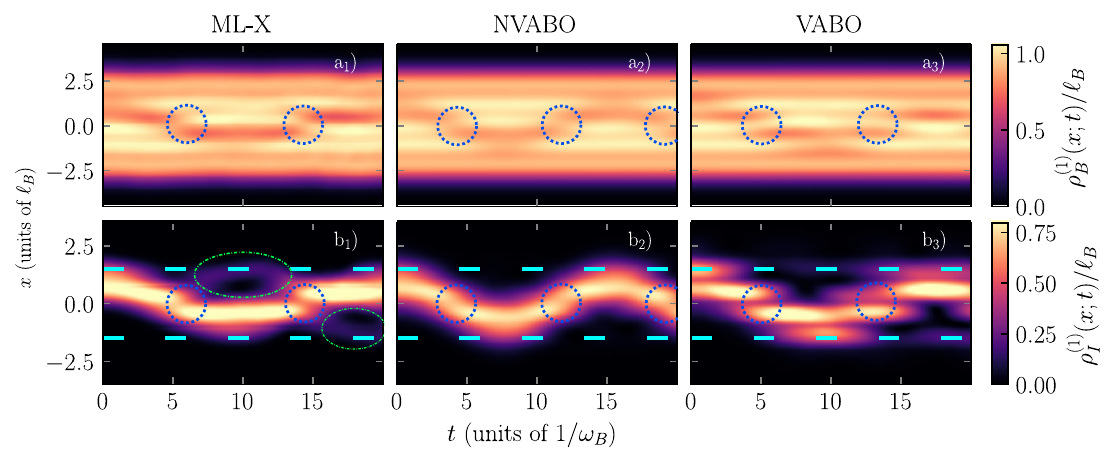}
    \caption{Spatiotemporal dependence of the one-body density of the bath $\rho_B^{(1)}(x;t)$ and impurity $\rho_I^{(1)}(x;t)$ species for an interacting system with $g=5.0\hbar\ell_B\omega_B$. The impurity mass $m_I=4m_B$ and trapping frequency $\omega_I=0.5\omega_B$ as well as the smaller shift $x_s=0.5\ell_B$ are considered. The different columns correspond to the distinct methods (see labels). For the impurity species, we include cyan dashed lines at $x = \pm(x_s+\sqrt{2}\ell_I)$ as a visual reference to the oscillation amplitude in the non-interacting case. \textcolor{black}{The dotted blue circles indicate the spatiotemporal regions where the impurity crosses $x_I = 0$, corresponding to the location of an avoided crossing. The green dotted circles in (b$_1$) highlight the reflected fraction of the impurity density by the avoided crossing at $x_I = 0$.}}
    \label{fig:one_body_density_g50_x05}
\end{figure*}
By introducing a weak interaction ($g=2.5\hbar\omega_B\ell_B$) between the bath and the single impurity, we present in Fig.~\ref{fig:one_body_density_g25_x05} the corresponding time evolutions of the densities $\rho^{(1)}_B(x;t)$ and $\rho^{(1)}_I(x;t)$. 
Starting with the numerically exact results from the ML-X method, we observe a similar dipole oscillation of the impurity, see Fig.~\ref{fig:one_body_density_g25_x05}(b$_1$), when compared to the non-interacting case, see Fig.~\ref{fig:one_body_density_g00}(b$_1$). However, the interaction introduces key differences from the non-interacting evolution. In particular, by comparing the interacting case of Fig.~\ref{fig:one_body_density_g25_x05}(a$_1$) and (b$_1$) to the non-interacting results of Fig.~\ref{fig:one_body_density_g00}(a$_1$) and (b$_1$) we observe that due to the bath-impurity repulsion the bath species develops a density valley at the position of the impurity species. 
Additionally, in the time instances that the impurity crosses $x_I \approx 0$, \textcolor{black}{which are marked by the dotted blue circles,} e.g. for $t \approx 4 \omega_B^{-1}, 12 \omega_B^{-1}$, its density maximum decreases. In addition in the elapsed time between these transits an impurity density accumulation is observed for $x \approx -0.5$ for $4 < \omega_B t < 10$ and $x \approx 0.5$ for $12 < \omega_B t < 18$, rendering the form of the impurity oscillation non sinusoidal in contrast to the non-interacting case, see Fig.~\ref{fig:one_body_density_g00}.
This effect is even more pronounced in the bath one-body density, leading to the pronounced formation of distinct minima in the spatial regions where the impurity accumulation occurs for \(x_B \approx 0.5~\ell_B\) and \(x_B \approx -0.5~\ell_B\), such that the bath depletion has a noticeably rectangular oscillation pattern. Another notable phenomenon for $\rho^{(1)}_I(x;t)$ is the reduction of its oscillation period, by comparing the oscillation pattern in Fig.~\ref{fig:one_body_density_g00}(b$_1$) with that in Fig.~\ref{fig:one_body_density_g25_x05}, we clearly see that the oscillation period is increased in the interacting case. Notice that both the ML-X and MCBO methods yield identical results, as both are numerically exact \textit{ab initio} approaches (not shown here for brevity).

To examine the degree of non-adiabatic effects that occur in the system we consider the simplest approximation provided by the NVABO, where all non-adibatic effects are completely neglected. As shown in Fig.~\ref{fig:one_body_density_g25_x05}(a$_2$), we observe that the structure of the bath one-body density is qualitatively similar to Fig.~\ref{fig:one_body_density_g25_x05}(a$_1$). However, the bath-density minimum caused by the interaction between bath and impurity species is less pronounced within NVABO, see Fig.~\ref{fig:one_body_density_g25_x05}(a$_2$). In the case of the impurity species, the density accumulation at \(x > 0\) and \(x < 0\) is still present but much less prominent, see Fig.~\ref{fig:one_body_density_g25_x05}(b$_2$). 
The oscillation period is shorter compared to the exact ML-X result, compare Fig.~\ref{fig:one_body_density_g25_x05}(b$_2$) to  Fig.~\ref{fig:one_body_density_g25_x05}(b$_1$). Hence, we conclude that beyond the adiabatic approximation, the Born-Huang contribution and non-adiabatic effects might be necessary to fully describe the prolongation of the oscillation period, which we investigate below by invoking the VABO approximation.

Indeed, the inclusion of the Born-Huang term (Fig.~\ref{fig:one_body_density_g25_x05}(a$_3$) and \ref{fig:one_body_density_g25_x05}(b$_3$)) results in a larger oscillation period of the impurity species compared to the NVABO approximation and comparable to the ML-X result. However, the spatial development of individual minima (at $x_I\approx 0$) and maxima (at $x_I = \pm 0.5$) of $\rho^{(1)}_{I}(x_I;t)$ are significantly enhanced (see Fig.~\ref{fig:one_body_density_g25_x05}(b$_3$)) in comparison to the ML-X result depicted in Fig.~\ref{fig:one_body_density_g25_x05}(b$_1$).
Thus, we can infer that the Born-Huang term, which accounts for changes in the kinetic energy of the bath, not only elongates the oscillation period of the impurity species, but also is the reason for the large fluctuations of the impurity density during its oscillation within the fermionic enviroment, leading to a density evolution reminiscent of transport within a multi-well setup~\cite{Cao_2011, Koutentakis2017}. 
However, since the importance of this effect is greatly overestimated within VABO we can infer that non-adiabatic transport to higher-lying PEC is important for correctly capturing the system dynamics even moderately away from the synthetic conical intersections, that as shown in Ref.~\cite{Becker2024}, occur in the limit $g\to\infty$.
\begin{figure*}
    \centering
    \includegraphics[width=0.9\linewidth]{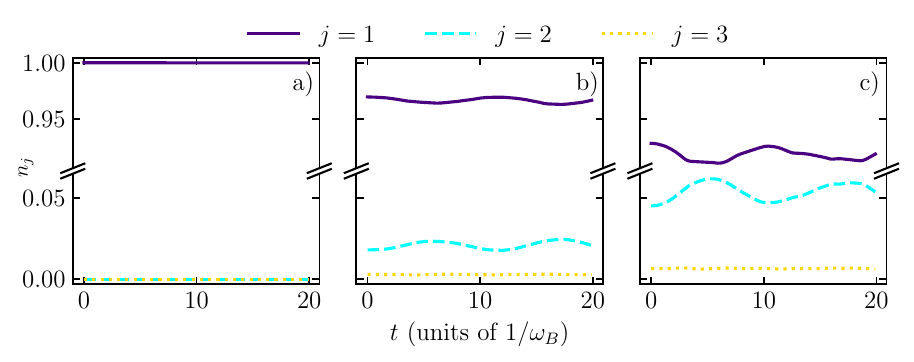}
    \caption{The total occupation of the first three PEC calculated via Eq. \eqref{eqn:pes_occ}. An initial shift $x_s=0.5\ell_B$, $m_I=4m_B$ and $\omega_I=0.5\omega_B$ are considered. Panel (a) corresponds to the non-interacting case, (b) is the weakly interacting case $g=2.5\hbar\ell_B\omega_B$ and lastly (c) the strong interaction case $g=5.0\hbar\ell_B\omega_B$. In all cases, $N_B = 5$.}
    \label{fig:sum_plot_occ_05}
\end{figure*}

Nevertheless, the analysis for \(g=2.5\hbar\ell_B\omega_B\) demonstrated that the approximate methods, namely the VABO and its non-variational form (NVABO), were useful to some extent in capturing some of the key aspects of the impurity and bath dynamics. These approaches provided a reasonably accurate description of the system’s density evolution, highlighting their applicability in the weakly interacting regime. However, as we now turn to the strongly interacting regime at \(g=5\hbar\ell_B\omega_B\), the reliability of these methods is expected to diminish, owing to the approach to the synthetic conical intersections at $g \to \infty$. 

To explore this possibility, we perform an one-body density analysis of the $g = 5~\hbar \omega_B\ell_B$ case in Fig.~\ref{fig:one_body_density_g50_x05}. Starting again with the one-body density of the bath, Fig.~\ref{fig:one_body_density_g50_x05}(a\(_1\)), and impurity species, Fig.~\ref{fig:one_body_density_g50_x05}(b\(_1\)), within ML-X. Similar to Fig.~\ref{fig:one_body_density_g25_x05}(a\(_1\)) and (b\(_1\)), we observe density minima in the bath one-body density (a\(_1\)) and again a minimum for $x_I\approx0$ (\textcolor{black}{see the blue dotted circles}) in the impurity one-body density. As expected, since these effects are a consequence of the interaction strength, it becomes more dominant for the strongly interacting case. In addition, the detected prolonging of the dipole oscillation period of the impurity species increases. An additional effect, hardly noticeable in the case of smaller interactions, compare Fig.~\ref{fig:one_body_density_g25_x05}(b$_1$) to~\ref{fig:one_body_density_g50_x05}(b$_1$), is that when the impurity species crosses the center of the harmonic trapping potential, for $g =5.0\hbar\omega_B\ell_B$ we detect a small part of the impurity species that is reflected at $x_I\approx 0$ and performs a mirror-symmetric oscillation \textcolor{black}{as indicated by the green dotted circles}. Furthermore, we observe that the impurity species dipole oscillation amplitude is significantly reduced when compared to the non-interacting case, compare Fig.~\ref{fig:one_body_density_g25_x05}(b$_1$) to Fig.~\ref{fig:one_body_density_g00}(b\(_1\)). 
The oscillation curves deviate from the initial sinusoidal profile observed in the dipole oscillation for the non-interacting case Fig.~\ref{fig:one_body_density_g00}(b$_1$) and possess an approximately rectangular shape.

Within the simplest approximation in the NVABO case, see Fig.~\ref{fig:one_body_density_g50_x05}(b$_2$), the impurity evolution is remarkably similar to the result of the NVABO approximation in the weakly interacting case ($g=2.5\hbar\omega_B\ell_B$) of Fig.~\ref{fig:one_body_density_g50_x05}(b$_2$). The main differences are relatively larger fluctuations of the $\rho^{(1)}_I(x_I;t)$ maximum, due to the stronger interaction between both species. Noticeably, the ML-X findings that set apart the weakly from the strongly interacting case explained above cannot be reproduced within NVABO. For instance, the oscillation period is only minimally extended by the stronger interaction, compare Fig.~\ref{fig:one_body_density_g50_x05}(b$_2$) to Fig.~\ref{fig:one_body_density_g25_x05}(b$_2$), significantly less than in the ML-X result. Also here the impurity oscillation pattern is very close to sinusoidal and the density depletion of the bath is not very different than for $g=2.5\hbar\omega_B\ell_B$ case, compare Fig.~\ref{fig:one_body_density_g50_x05}(b$_1$) to Fig.~\ref{fig:one_body_density_g25_x05}(b$_1$).

The density evolution when taking into account the Born-Huang term within the VABO approximation is depicted in Fig.~\ref{fig:one_body_density_g50_x05}(a$_3$) and \ref{fig:one_body_density_g50_x05}(b$_3$). In particular, here, in addition to the impurity density maxima at \(x_I \approx \pm 0.5~\ell_B\) another set of maxima appears around \(x_I \approx \pm 1.0~\ell_B\), see (Fig.~\ref{fig:one_body_density_g50_x05}(a$_3$)). This again demonstrates the tendency of the impurity within VABO to behave similarly to a multi-well setup, a behavior that is again much more overestimated within this approach that in the numerically exact one, compare Fig.~\ref{fig:one_body_density_g50_x05}(a$_2$) and \ref{fig:one_body_density_g50_x05}(c$_2$).
Additional differences to the exact case are that the amplitude of the oscillation of the impurity species remains equally wide to the non-interacting case within VABO and the oscillation pattern remains characteristically sinusoidal. However, with regard to the oscillation period duration, the VABO approximation differs significantly from the NVABO approximation and is very close to the ML-X result.
\subsubsection{Analysis in terms of PEC}
\begin{figure*}
    \centering
\includegraphics[width=0.9\linewidth]{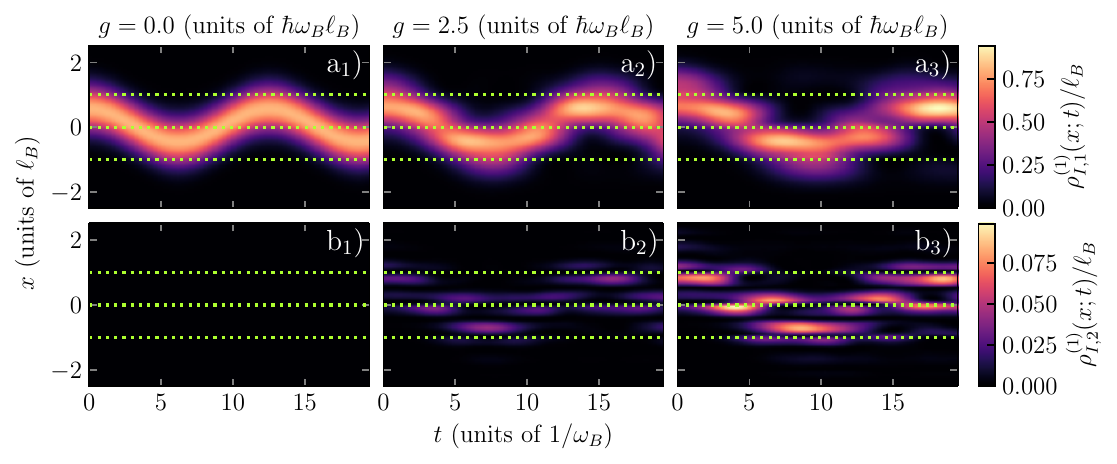}
    \caption{Time dependence of the PEC-resolved one-body density of the impurity, $\rho_{I,n}^{(1)}(x;t)$, corresponding to the $n$-th PEC and for varying interaction strengths (see label). The impurity mass is set to $m_I = 4m_B$ and the trapping frequency to $\omega_I = 0.5\omega_B$, with a smaller shift of $x_s = 0.5\ell_B$ taken into account. The green dashed lines indicate the avoided crossings between the ground-state and first-excited PEC, as described in Eq.~\eqref{eqn:potential_energy_surfaces}.
}
    \label{fig:pec_occ_xs_05}
\end{figure*}

The above confirms our initial expectation that the interaction strength emerges as a decisive factor in the non-adiabaticity of our system. 
In the interacting case, the adiabatic approximations capture the qualitative behavior of the system to some extent. However, a more detailed examination reveals important qualitative differences to the case where the non-adiabatic couplings are taken into account. While the qualitative behavior of the system is not very sensitive to the interaction strength, on the quantitative level the intensity of the observed non-adiabatic phenomena is greatly amplified as the synthetic conical intersections at $g \to \infty$ are approached. 
Therefore, incorporating non-adiabatic corrections within the framework of MCBO or another numerically exact approach, such as ML-X, is necessary to achieve a quantitatively accurate description of these amplified effects. 
In addition, even when restricting our focus to the adiabatic approximation, the Born-Huang correction \(V_{11}^{\rm ren}(x_I)\) proves to be a clearly relevant factor, especially in the regime of strong interactions. This underscores that the response of the bath in terms of its kinetic energy is essential to get the full picture of the system dynamics.

Building on the insights from the previous section, where we have highlighted the crucial role of non-adiabaticity in the dynamics of an interacting bath-impurity system. We now shift our focus to a more detailed exploration of the system’s behavior through the lens of the PEC occupations. This step allows us to connect the non-adiabatic effects to the observed dynamics, offering a clearer picture of how each PEC influences the overall system. By investigating the contributions of the individual PEC, we are able to directly connect them to the PEC avoided crossing associated to the PJTE~\cite{Becker2024}.
\begin{figure*}
    \centering
    \hspace*{-2cm}\includegraphics[width=1.2\linewidth]{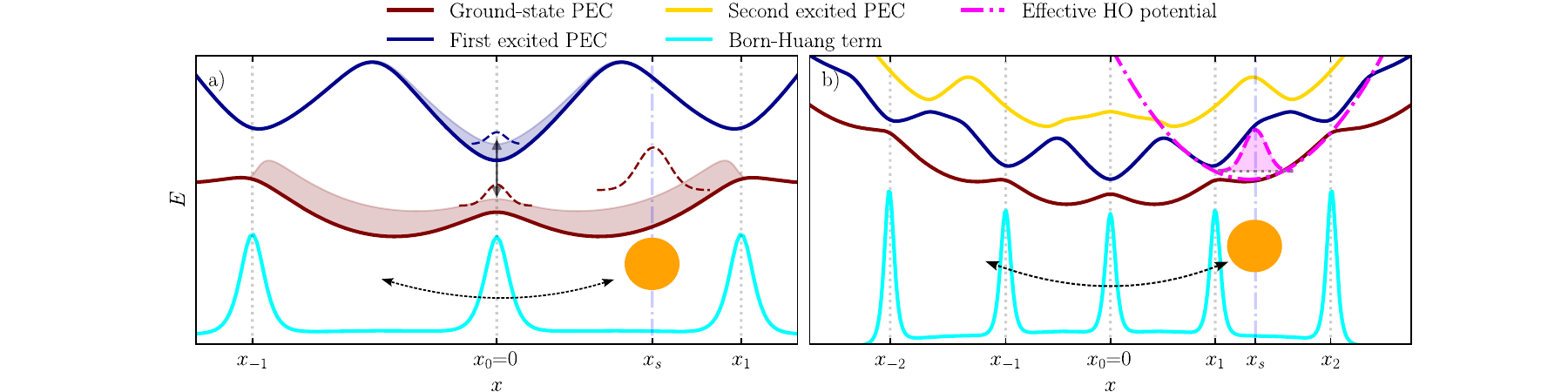}
    \caption{\textcolor{black}{Impurity dynamics on the PECs are shown in the two regimes of impurity displacement. In both subfigures the Born-Huang term  \(V_{11}^{\rm ren}(x_I)\), which scales with the interaction strength \(g\), is shown with a constant vertical offset for better comparison with the PECs.
    The avoided crossings, marked by vertical lines at positions $x_i$ for $i=-2,\dots,2$, coincide with the maximum peaks of the Born-Huang term  \(V_{11}^{\rm ren}(x_I)\), underscoring the importance of non-adiabatic coupling and interaction strength in shaping the impurity dynamics.
(a) Small shift regime $x_s = 0.5 \ell_B$: The impurity exhibits quasi-adiabatic transport, except near the avoided crossing $x_0=0$, where density is exchanged between the lowest and first excited PEC.} \textcolor{black}{The PECs are derived for an interaction strength of \( g = 5.0\,\hbar\ell_B\omega_B \), impurity mass \( m_I = 4m_B \), and impurity trapping frequency \( \omega_I = 0.5\omega_B \). To highlight the avoided crossings, the PECs are scaled and shifted linearly. This adjustment enhances visual contrast without altering the underlying physical trends.
The width of the faintly colored regions on the PECs indicates the relative contributions to each PEC.}
\textcolor{black}{(b) Large shift regime $x_s = 1.5 \ell_B$: The impurity becomes self-trapped in an effective harmonic potential in the outer region of the lowest PEC, exhibiting quasi-diabatic behavior. As a result, only a small fraction of the impurity density can tunnel out of this potential.}}
\label{fig:impurity_dynamics}
\end{figure*}

To solidify this connection we first analyze the occupation of each PEC $j$,
\begin{equation}
    n_j(t) \equiv \langle \Psi (t)|\hat{\mathcal{P}}_j|\Psi (t) \rangle = \int\mathrm{d}x_I~|\Psi_{j,I}(x_I;t)|^2.
    \label{eqn:pes_occ}
\end{equation}
for details on the derivation of this quantity, see Appendix \ref{app:Derivations in Second Quantization Formalism}. We show $n_j(t)$ as a function of time in Fig.~\ref{fig:sum_plot_occ_05} for the dominating lowest three PECs. Here Fig.~\ref{fig:sum_plot_occ_05}(a) refers to the simple non-interacting case, where no excited PEC are occupied and the adiabatic approximation is sufficient to describe the physics. As worked out in our extensive ground-state analysis \cite{Becker2024}, turning on the interaction leads to beyond Born-Oppenheimer physics even in the static case, because of the PJTE. 
Indeed, in the weakly interacting case, $g=2.5\hbar\omega_B\ell_B$, of Fig.~\ref{fig:sum_plot_occ_05}(b) both the first and second PEC are populated already for the initial state, $t = 0$. During the time evolution, we observe additional population transfer, predominantly between the ground $j = 1$ and the first excited, $j = 2$, PEC. This verifies our claims of non-adiabatic effects being important during the dynamics of the system. For $g = 5.0 \hbar\omega_B\ell_B$ depicted in Fig.~\ref{fig:sum_plot_occ_05}(c), the above results are more pronounced. In particular, we observe that the occupation of the $j = 1$ is more suppressed even for $t = 0$, as expected due to the approach to the synthetic conical intersections for $g \to \infty$. In addition, during the oscillation dynamics of the impurity, the transfer of the PEC occupation is amplified compared to that of the weakly interacting case. The resulting $n_j(t)$ oscillations show dephasing as the second oscillation period has a significantly longer duration than the first. 

However, a deeper look at the population transfer oscillations reveals an open point in our analysis.
In particular, inspecting Fig.~\ref{fig:sum_plot_occ_05}(b) and~\ref{fig:sum_plot_occ_05}(c), we observe that quite peculiarly the frequency of the population transfer does not match the frequency of the dipole oscillations of the impurity, see Fig.~\ref{fig:one_body_density_g25_x05}(b$_1$) and~\ref{fig:one_body_density_g50_x05}(b$_1$) respectively.
More specifically, by inspecting Fig.~\ref{fig:sum_plot_occ_05}(b), in the case of $g = 2.5~\hbar \omega_B \ell_B$ we observe that the first minimum of $n_1(t)$ occurs at $t \approx 6.2 \omega_B^{-1}$, where $\rho^{(1)}_{I}(x;t)$, see Fig.~\ref{fig:one_body_density_g25_x05}(b), is close to its minimum negative displacement, while the second maximum is at $t \approx 12.1\omega_B^{-1}$, when $\rho^{(1)}_I (x;t)$ crosses $x = 0$. Finally, the second minimum visible in Fig.~\ref{fig:one_body_density_g25_x05}(b) corresponds to $t = 16.6~\omega_B^{-1}$, where the impurity travels from its maximum positive displacement toward $x = 0$. Similar patterns are also identified for the strongly interacting $g = 5~\hbar \omega_B \ell_B$ state. It can then be claimed that there is little correlation of PEC transfer processes and the synthetic conical intersections and PJTE given the not straightforward relation of the PEC population transfer and the location where it occurs. \textcolor{black}{To illustrate the discussed impurity dynamics, we provide a schematic representation in Fig.~\ref{fig:impurity_dynamics}(a), focusing on the relevant PECs and the contribution of the Born-Huang term \( V_{\rm ren}^{11}(x_I) \). By applying a linear shift to the Born-Huang term, we can qualitatively compare its spatial behavior with the PECs and highlight that its maxima coincide with the avoided crossings. Additionally, the PECs are rescaled to emphasize the appearance of these avoided crossings. As previously derived, this illustrates how the impurity undergoes quasi-diabatic transport governed by the structure of the PECs, except near the avoided crossing at \( x_0 \approx 0 \), where an exchange of contribution occurs between the ground and first excited states.} However, by further analysis of the occupied impurity configurations within each PEC we can reveal the hidden relationship between these transfer processes and the PJTE.

As a first step to address this issue we examine the impurity dynamics within the two energetically lowest PEC. To achieve this we analyze the impurity density evolution, $\rho^{(1)}_{I, j}(x;t)$ for the PEC, $j = 1,2$, by exploiting the fact that within MCBO the total impurity density reads
\begin{equation}
    \rho^{(1)}_I(x_I;t) = \sum_{j = 1}^{M} \rho^{(1)}_{I, j} (x_I;t) = \sum_{j = 1}^{M} |\Psi_{j,I}(x_I;t)|^2.
\end{equation}
Here we note again that the results of MCBO are almost identical to ML-X, since both methods are numerically exact and therefore employing the fact that $\rho^{(1)}_{I, j} (x_I;t)$ is much easier to calculate within the former approach is legitimate. The results of our analysis are shown in Fig.~\ref{fig:pec_occ_xs_05}. The first feature that this analysis clearly demonstrates is that the density of the impurity within the $j = 1$ PEC is almost identical to the VABO approximation, compare Fig.~\ref{fig:pec_occ_xs_05}(a$_2$) to  Fig.~\ref{fig:one_body_density_g25_x05}(b$_3$) and Fig.~\ref{fig:pec_occ_xs_05}(a$_3$) to  Fig.~\ref{fig:one_body_density_g50_x05}(b$_3$). This justifies our argument in the previous section that the differences between VABO and ML-X are attributed to occupation of higher lying PEC. Turning our attention to the dynamics within the second PEC, we observe that the occupation of this PEC in the initial state is associated to states close to the avoided crossing of the PEC at $x \approx -1~\ell_B$, see Fig.~\ref{fig:setup_and_potential_energy_surfaces}(c), note that the positions of the avoided crossings have been marked by horizontal dashed lines in Fig.~\ref{fig:pec_occ_xs_05}(b$_2$) for convenience. Later in the dynamics, as the impurity crosses $x = 0$ for $t \approx 4~\omega_B^{-1}$ we observe that $\rho^{(1)}_{I,2}(x;t)$ is transferred from the vicinity of the $x \approx 1~\ell_B$ to the $x = 0$ avoided crossing. This pattern continues with the occupation of the second PEC being transferred to the avoided crossing closest to where $\rho^{(1)}_{I, 1}(x;t)$ lies during its dipole oscillations, compare Fig.~\ref{fig:pec_occ_xs_05}(a$_2$) and Fig.~\ref{fig:pec_occ_xs_05}(b$_2$). Nevertheless, we can see that a portion of $\rho^{(1)}_{I,2}(x;t)$ persists in the vicinity of $x = 0$ avoided crossing throughout the time evolution. This density fraction shows oscillations between the $x > 0$ and $x < 0$ part of the avoided crossing with a frequency that is significantly faster from the dipole oscillation frequency of the impurity. A similar pattern is also observed for Fig.~\ref{fig:pec_occ_xs_05}(a$_3$) and~\ref{fig:pec_occ_xs_05}(b$_3$), but the population of the second PEC is much more pronounced.
\begin{figure*}
    \centering
    \includegraphics[width=1.0\linewidth]{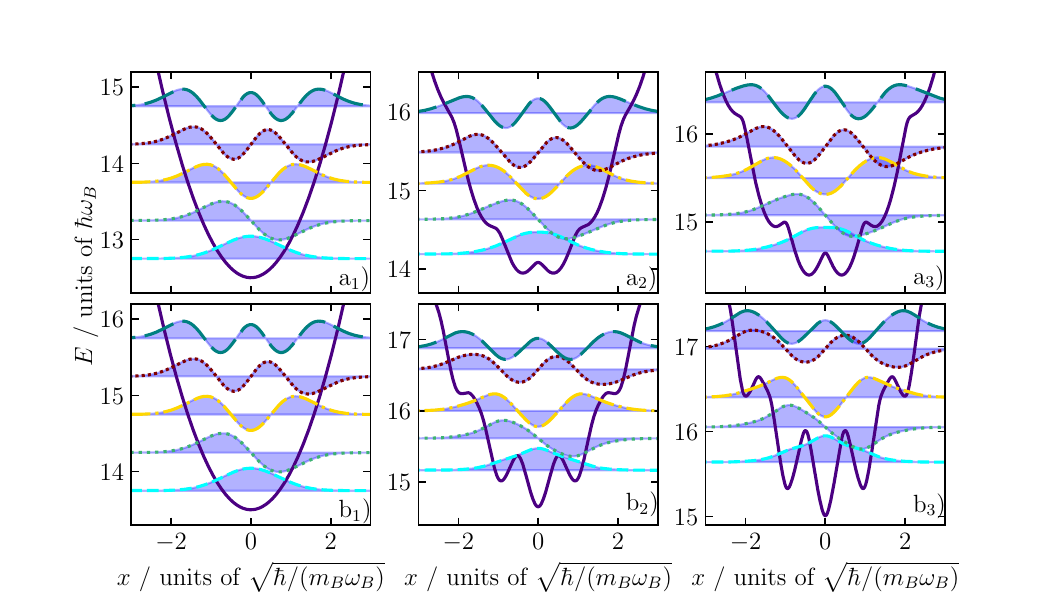}
    \caption{The effective potential of Eq.~\eqref{effeective-potential-sp} combining the harmonic trapping potential with the PEC $\varepsilon_j(x)$. The dashed and dotted lines show its eigenfunctions, scaled by a factor of two and offset by their corresponding eigenenergy, i.e. $2\psi_j^{(m)}(x)+E_j^{(m)}$. The upper panels (a$_i$) refer to the lowest PEC $j=1$, the lower panel (b$_i$) show the behavior of the first excited PEC $j=2$. The index $i=1,2,3$ refers to the interaction strengths $g=0.0$, $g=2.5\hbar\omega_B\ell_B$ and $g=5.0\hbar\omega_B\ell_B$.}
    \label{fig:lowest_PES_eigenstates}
\end{figure*}

\begin{figure*}
    \centering
    \hspace*{-1cm} 
    \includegraphics[width=1.1\linewidth]{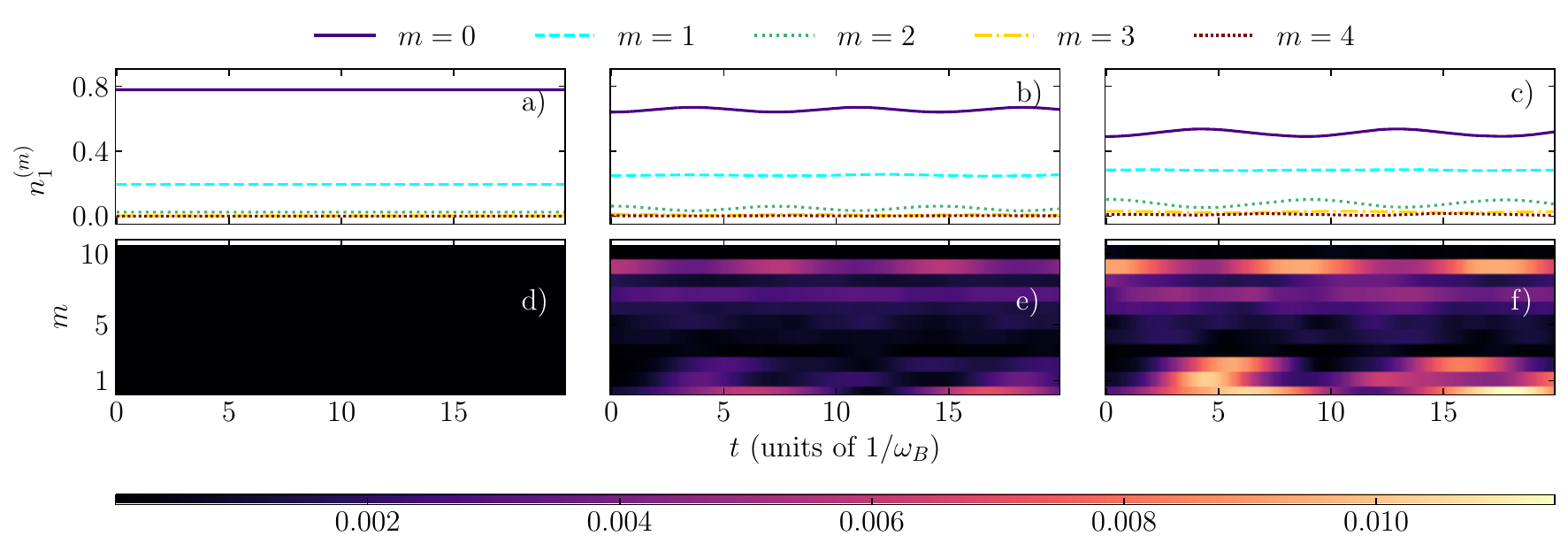}
    \caption{Time-evolution of the population $n_j^{(m)}$ of the eigenstates of (a), (b), (c) the ground-state PEC $j=1$ and (d), (e), (f) the first excited $j=2$ for $m=0,\cdots,10$, see Eq.~\eqref{eqn:states_m_pes_j_ev}. In all cases a shift $x_s=0.5$ is considered. The interaction strength is (a) and (d) $g=0.0$,  (b) and (e) $g=2.5$, and (c) and (f) $g=5.0$.}
    \label{fig:2D_plot_occ_05}
\end{figure*}
The above analysis provides important information on the dynamics of the system with respect to the PEC occupation. First, it is obvious that the avoided crossings of the PEC are prominently involved as substantial $\rho^{(1)}_{I, 2}(x;t)$ is observed only close to these regions. It can be argued that the frequency mismatch between the population transfer to the impurity dipole oscillation dynamics, observed in Fig.~\ref{fig:sum_plot_occ_05} can be attributed to the additional fast dynamics related to the tunneling of the one-body density of the second PEC in the region $x = 0$ observed in Fig.~\ref{fig:pec_occ_xs_05}(b$_2$) for weak interactions ($g=2.5\hbar\omega_B\ell_B$) and also for strong interactions ($g=5.0\hbar\omega_B\ell_B$) in Fig.~\ref{fig:pec_occ_xs_05}(b$_3$). 
However, what remains to be clarified is what processes lead to the generation of this frequency and the observed phenomenology. To achieve this, we delve deeper into the intra-PEC dynamics by analyzing the occupation of different spatial configurations within each PEC.

For this purpose we consider an effective one-body Hamiltonian to describe the different impurity-species spatial configurations
\begin{equation}
    \begin{split}
        \hat{H}_{\rm PEC}^{j}=-\frac{\hbar^2}{2m_I}\left(\frac{\partial}{\partial x_I}\right)^2+\frac{1}{2}m_B\omega_I^2 x_I^2+\varepsilon_j(x_I).
    \end{split}
    \label{effeective-potential-sp}
\end{equation}
This effective potential consists of the kinetic term, the harmonic trapping confinement of the impurity and the energy of the $j$-th PEC, $\varepsilon_j(x)$ given by  \eqref{eqn:potential_energy_surfaces} for a fixed bath-impurity interaction strength $g$. In the spirit of the adiabatic Born-Oppenheimer approximation, this can be considered as the equation of motion of the slow and heavy degree of freedom.
This system effectively describes the impurity's behavior, with the PEC $\varepsilon_j(x)$ acting as an effective potential that accounts for the contact interaction with the bath species. 
We diagonalize Eq.~\eqref{effeective-potential-sp} in the context of the static Schr\"odinger equation
\begin{equation}
    \hat{H}_{\rm PEC}^{j} \phi_{j,m}(x) = E_{j,m}\phi_{j,m}(x),
    \label{eqn:one_body_ham}
\end{equation}
with its corresponding eigenfunctions, $\phi_{j,m}(x)$, and eigenenergies, $E_{j,m}$, shown in Fig.~\ref{fig:lowest_PES_eigenstates}. For the non-interacting case and for both the ground-state PEC [Fig.~\ref{fig:lowest_PES_eigenstates}(a$_1$)], and the first excited PEC [Fig.~\ref{fig:lowest_PES_eigenstates}(b$_1$)], we identify an effective harmonic potential ($\sim x^2$), which is the same for both the lowest and the higher-lying PECs, \textcolor{black}{which remain identical for both levels apart from a constant shift, reflecting the spatially uniform nature of the PECs in the non-interacting regime}. Introducing interactions ($g=2.5$ in Fig.~\ref{fig:lowest_PES_eigenstates}(a$_2$) and \ref{fig:lowest_PES_eigenstates}(b$_2$)) between the bath and impurity species results in a double-well potential for the lowest PEC, whereas the first excited state exhibits a triple-well structure. These features become more pronounced in the strongly interacting regime ($g=5.0\hbar\omega_B\ell_B$ Fig.~\ref{fig:lowest_PES_eigenstates}(a$_3$) and (b$_3$)), with the triple-well potential of the $j = 2$ PEC emerging particularly clearly. \textcolor{black}{When examining our effective potential, we observe in Fig.~\ref{fig:lowest_PES_eigenstates}(b$_2$), (a$_3$), and (b$_3$) that, in addition to the triple-well structure, additional potential wells are present. However, these appear at such high energies that they do not significantly contribute to our analysis.}

This structure is of particular interest. The triple-well potential can capture impurity density when the impurity crosses one of its wells. This generates two pathways for the dynamics, first the diabatic excitation of the impurity to an excited PEC or vice versa and second the intra-PEC tunneling dynamics among the wells, which can be used to explain the findings of Fig.~\ref{fig:pec_occ_xs_05}.

To examine this issue further, we now focus on the occupation of the individual states of the first two PECs. 
Hence, we calculate the occupation of the individual states $m$ of the corresponding PEC $j$
\begin{equation}
n_j^{(m)}=\left|\int\mathrm{d}x_I\Psi_{j,I}^{*}(x_I)\phi_{m,j}(x_I)\right|^2,
\label{eqn:states_m_pes_j_ev}
\end{equation}
details on the derivation of this expression is found in Appendix \ref{app:Derivations in Second Quantization Formalism}.
Considering the non-interacting case, Fig.~\ref{fig:2D_plot_occ_05}(a) reveals that the impurity resides in a coherent state with populations consistent with \(x_s = 0.5 \ell_B\). As previously mentioned the second PEC is completely depopulated in this case due to the absence of interactions, see Fig.~\ref{fig:2D_plot_occ_05}(d).
Introducing interactions with strength \( g=2.5\hbar\omega_B\ell_B \) fundamentally alters the system's behavior. By contrasting Fig.~\ref{fig:2D_plot_occ_05}(b) to the non-interacting scenario shown in Fig.~\ref{fig:2D_plot_occ_05}(a) and~\ref{fig:2D_plot_occ_05}(d), we observe a shift of the initial population of the states of the first PEC, especially apparent is the reduction of the corresponding ground states for $m = 0$. In addition, in contrast to the constant in time populations of Fig.~\ref{fig:2D_plot_occ_05}(a), the occupations of the different $m$ contributions in the interacting case of Fig.~\ref{fig:2D_plot_occ_05}(b), are not constant in time but they rather fluctuate with double the frequency of the impurity's dipole oscillation. This behavior can be attributed to the impact of the Born-Huang term as it can be captured well by VABO (not shown here for brevity).
\begin{figure*}
    \centering
    \includegraphics[width=0.9\linewidth]{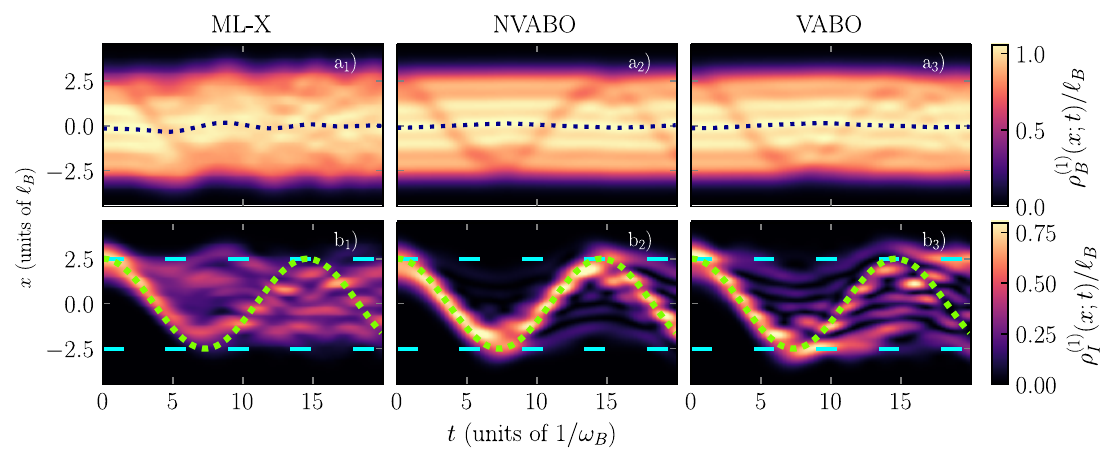}
    \caption{Spatiotemporal dependence of the one-body density of the bath $\rho_B^{(1)}(x;t)$ and impurity $\rho_I^{(1)}(x;t)$ species for an interacting system with $g=2.5\hbar\ell_B\omega_B$. The impurity mass $m_I=4m_B$ and trapping frequency $\omega_I=0.5\omega_B$ as well as the larger shift $x_s=1.5\ell_B$ are considered. The different columns correspond to the distinct methods (see labels). For the impurity species, we include cyan dashed lines at $x = \pm(x_s+\sqrt{2}\ell_I)$ as a visual reference to the oscillation amplitude in the non-interacting case. \textcolor{black}{The blue dotted line in the upper panels (a$_i$) indicates the average position of the bath density $\langle x_B(t) \rangle$. The green dotted line in the impurity density panel (b$_i$) gives as a guide to the eye the dipole oscillation in the NVABO case.}}
    \label{fig:one_body_density_g25_xs15}
\end{figure*} 

Turning our attention to the second $j = 2$ PEC we observe the population of the higher-lying \( m=9 \) state of the first excited PEC and to a lesser extent \( m = 7 \), see Fig.~\ref{fig:2D_plot_occ_05}(e). The above demonstrate that the initial state is affected by non-adiabatic couplings in agreement to \cite{Becker2024}. The involvement of such highly excited states of the $j = 2$ state is a remarkable consequence of the Born-Huang term and a manifestation of the influence of the synthetic conical intersection at $g \to \infty$ to the finite $g$ dynamics. Indeed, given the eigenstates of the effective potential Fig.~\ref{fig:lowest_PES_eigenstates}(b$_2$), one would expect that the first three states would be involved in the dynamics which represent the triple-well, however, this is not the case since close to the minima of the effective potential the energy correction stemming from the Born-Huang term becomes very large as does the non adiabatic coupling, $A_{1,2}(x_I)$ of Eq.~\eqref{eqn:schroedinger-time-dendent}. This shows that the adiabatic-basis that MCBO employs is not a good basis for describing the system, a well-known fact in molecular physics where approaches based on quasi-diabatic potentials \cite{Pacher1988, Pacher1989, Baer2006, Hummel2023} are preferred to treat the dynamics of a system close to a conical intersection.
During time evolution we observe that the occupations of these highly excited states fluctuate with double the frequency of the dipole oscillation, therefore we can conclude that this initial excited population essentially performs coherent oscillations within the excited PEC.
An additional effect appears at the time instances that the impurity crosses the $x = 0$ point, more precisely at $t \approx 4 \omega_B$ and $t \approx 12 \omega_B$, we detect a population transfer to the \( m=0 \) state of the first excited PEC. This is the fingerprint of $A_{1,2}(x_I)$ in the dynamics, reinforcing the pivotal role of avoided crossings in governing the system’s dynamical evolution. This excitation also results in dynamical transport\footnote{Formally this is overbarrier dynamics since the potential cannot localize states within the wells, see Fig.~\ref{fig:lowest_PES_eigenstates}(b$_2$)} from $m = 0$ to $m = 1, 2$ states is observed corresponding to redistribution of the impurity density among the wells, see Fig.~\ref{fig:lowest_PES_eigenstates}(b$_2$).
Based on this we can explain the fast oscillations observed close to $x = 0$ in Fig.~\ref{fig:pec_occ_xs_05}(b$_2$) as the results of the simultaneous tunneling dynamics of the initial excited $j = 2$ PEC density and the non-adiabatic transfer of impurity density leading to its trapping to the low-lying effective potential states.

As the interaction strength increases to \( g=5.0\hbar\omega_B\ell_B \), the trends identified in the weakly interacting regime become even more pronounced, as depicted in Fig.~\ref{fig:2D_plot_occ_05}(c) and (f). Specifically, transitions and occupations of the higher-lying states of the first PEC, most notably for \( m=7 \) and \( m=9 \), are further enhanced. This is accompanied by a concurrent suppression of the \( m=0 \) ground-state occupation within the first PEC, highlighting the growing influence of PJTE induced phenomena as the synthetic conical intersections at $g \to \infty$ are approached.

In our previous work~\cite{Becker2024}, we clearly established in our beyond Born-Oppenheimer investigation the association between avoided crossings and the PJTE. The present analysis, focusing in the beginning on the case of a small displacement, already demonstrates that the impurity dynamics is prominently affected by the PJTE.

\subsection{Role of interaction for the larger shift}

\begin{figure*}
    \centering
    \includegraphics[width=0.9\linewidth]{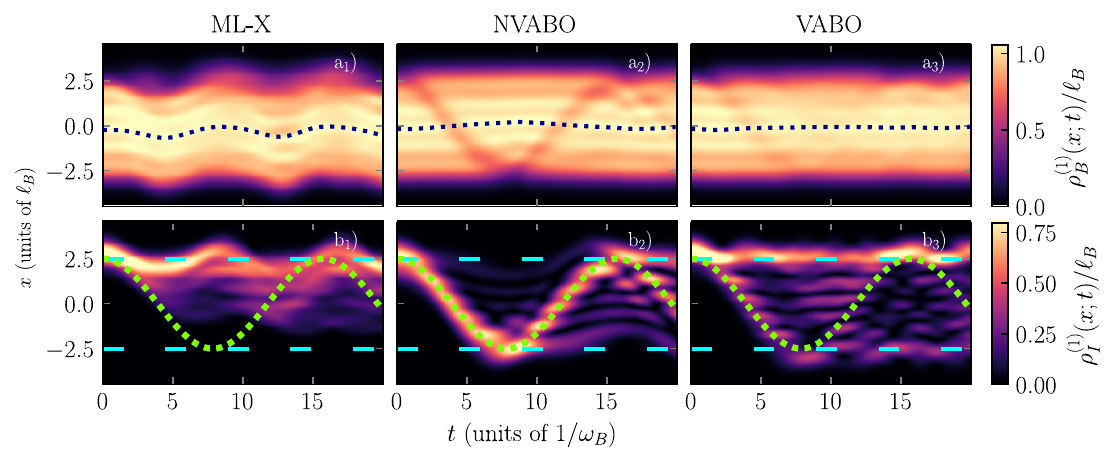}
    \caption{Spatiotemporal dependence of the one-body density of the bath $\rho_B^{(1)}(x;t)$ and impurity $\rho_I^{(1)}(x;t)$ species for an interacting system with $g=5.0\hbar\ell_B\omega_B$. The impurity mass $m_I=4m_B$ and trapping frequency $\omega_I=0.5\omega_B$ as well as the larger shift $x_s=1.5\ell_B$ are considered. The different columns correspond to the distinct methods (see labels). For the impurity species, we include cyan dashed lines at $x = \pm(x_s+\sqrt{2}\ell_I)$ as a visual reference to the oscillation amplitude in the non-interacting case. The blue dotted line in the upper panels (a$_i$) indicates the average position of the bath density $\langle x_B(t) \rangle$. The green dotted line in the impurity density panel (b$_i$) gives as a guide to the eye the dipole oscillation in the NVABO case.}
    \label{fig:one_body_density_g50_xs15}
\end{figure*}

Given that non-adiabatic effects are already manifest for $x_s = 0.5~\ell_B$ it is interesting to examine how our findings modify when additional avoided crossings, induced by the PJTE, come into play, as the initial shift, $x_s$, increases. As we will elaborate below an adjustment to $x_s = 1.5~\ell_B$ leads to a pronounced impact of non-adiabatic phenomena on the system dynamics, providing a deeper insight into how non-adiabatic effects govern the evolution of the system.

\subsubsection{Density analysis}
Analogously to Sec.~\ref{density_analysis_small}, we begin by examining the weaker interaction, laying the groundwork for analyzing the bath and impurity one-body density before progressing to stronger coupling regimes.
Hence, we consider in Fig.~\ref{fig:one_body_density_g25_xs15} the interaction strength $g=2.5\hbar\omega_B\ell_B$. Here, Fig.~\ref{fig:one_body_density_g25_xs15}(a\(_1\)) illustrates the one-body density of the bath, while Fig.~\ref{fig:one_body_density_g25_xs15}(b\(_1\)) presents the corresponding impurity one-body density, both derived from the numerically exact ML-X results. First, notice that as a result of the bath impurity interaction the initial state of the impurity for $g > 0$ is displaced with respect to the $g = 0$ case, compare the initial position of the density relative to the horizontal dashed lines in Fig.~\ref{fig:one_body_density_g25_xs15}(b$_1$) and Fig.~\ref{fig:one_body_density_g00}(b$_2$). The coherent dipole oscillation of the impurity species, evident in the non-interacting case for the complete time evolution, is only faintly visible up to \( t \approx 7.5 / \omega_B \) for $g = 2.5~\hbar \omega_B\ell_B$. In the latter case it becomes apparent that a significant portion of the impurity density is unable to transmit through the bath and it is instead getting trapped within its spatial extent. This can be attributed to the scattering of the impurity with its bath leading to momentum transfer among the species. 
Beyond $t \approx 7.5 \omega_B^{-1}$, $\rho^{(1)}_I(x;t)$ gets redistributed within its fermionic environment possessing substantial spatial modulations. In particular, it exhibits prominent interference-like peaks and troughs along the trajectory that a coherent state would have followed. 
While the bath density for $t < 7.5 \omega_B^{-1}$ shows a density depletion along the path of the coherent state, at later times it shows a highly fluctuating character, as shown in Fig.~\ref{fig:one_body_density_g25_xs15}(a\(_1\)) \textcolor{black}{and is represented by the oscillating average mean position of the bath species $\langle x_B(t) \rangle= {\int \mathrm{d}x~x\rho_B^{(1)}(x,t)}/{\int \mathrm{d}x~\rho_B^{(1)}(x,t)}$, depicted by the dotted blue line}. This behavior can also be attributed to the momentum exchange among the bath and impurity particles mentioned above.

Next, we evaluate to what extent this behavior is captured by the adiabatic approximation. Focusing on the simpler NVABO approximation presented in Fig.~\ref{fig:one_body_density_g25_xs15}(a\(_2\)) and (b\(_2\)), the dipole oscillation of the impurity species remains apparent in the impurity one-body density throughout the time evolution. The bath density exhibits a corresponding density depletion due to the bath-impurity repulsion, see Fig.~\ref{fig:one_body_density_g25_xs15}(a\(_2\)). In the impurity one-body density, a time-dependent modulation of the peak impurity density during its oscillation becomes noticeable after the first half-period, particularly visible in Fig.~\ref{fig:one_body_density_g25_xs15}(b\(_2\)) around \( x_I \approx 0 \) and \( x_I \approx \pm 1 \ell_B \). Moreover, after the first period of oscillation, a small fraction of the impurity density fails to penetrate back into the bath density, see Fig.~ \ref{fig:one_body_density_g25_xs15}(b\(_2\)) for $t \approx 15 \omega_B^{-1}$, leading to a small density fraction lying outside the fermionic environment. 
\begin{figure*}
    \centering
    \includegraphics[width=0.8\linewidth]{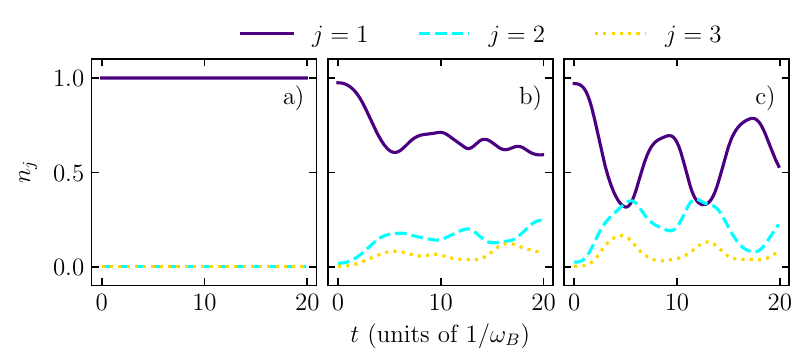}
    \caption{The total occupation of the first three PEC calculated via Eq. \eqref{eqn:pes_occ}. An initial shift $x_s=1.5\ell_B$, $m_I=4m_B$ and $\omega_I=0.5\omega_B$ are considered. Panel (a) corresponds to the non-interacting case, (b) is the weakly interacting case $g=2.5\hbar\ell_B\omega_B$ and lastly (c) the strong interaction case $g=5.0\hbar\ell_B\omega_B$. In all cases, $N_B = 5$.}
    \label{fig:sum_plot_occ_15}
\end{figure*}

When comparing the dipole oscillation of the impurity species in Fig.~\ref{fig:one_body_density_g25_xs15}(b\(_2\)) to the non-interacting case (Fig.~\ref{fig:one_body_density_g00}(b\(_2\))), we observe a reduction in the oscillation period, underscoring the influence of interactions within the NVABO approximation, however this reduction is greatly underestimated when compared to ML-X. This is similar to the case of the smaller shift (Fig.~\ref{fig:one_body_density_g25_x05}(b\(_2\))) at $x_s=0.5\ell_B$, where the dipole oscillation period remains nearly unchanged compared to the underlying case. This suggests that within the NVABO approximation, the dipole oscillation frequency is largely insensitive to the interaction. Further supporting our argument which attributes this frequency modulation to the Born-Huang term associated to the bath momentum.  

Turning our attention to the Born-Huang correction within the VABO approximation, Fig.~\ref{fig:one_body_density_g25_xs15}(a\(_3\)) shows a similar behavior for the bath one-body density to the NVABO approximation. However, in Fig.~\ref{fig:one_body_density_g25_xs15}(b\(_3\)), it is evident that the Born-Huang term prevents the impurity species from penetrating the bath species after the first half period, $t > 7.5\omega_B^{-1}$\textcolor{black}{, as evidenced by the loss of agreement with the dipole oscillation predicted by the NVABO approximation}. The pronounced dipole oscillation seen in the NVABO case gets distorted for larger times, as the impurity density fragments into smaller components that exhibit a fluctuating density pattern. Even in the VABO case, however, the fraction of the impurity density that deviates from the dipole oscillation pattern is much smaller when compared to the numerically exact result in Fig.~\ref{fig:one_body_density_g25_xs15}(b\(_1\)). In agreement to our previous arguments, the period of the dipole oscillation is consistent with the ML-X result of Fig.~\ref{fig:one_body_density_g00}(b\(_1\)) and significantly longer than NVABO, see Fig.~\ref{fig:one_body_density_g00}(b\(_2\)).

Already by examining this case we observe a dramatic shift in the phenomena emerging in the dynamics as the non-adiabatic effect associated to the excitation of the bath by the motion of the impurity become more apparent and thus both adiabatic approximations fail to capture some of the essential features of the dynamics. By considering even larger interaction values we can show the complete breakdown of the adiabatic dynamics.

Indeed, by turning our attention to the strongly interacting scenario (\(g = 5.0 \hbar \omega_B \ell_B\)) we observe a dramatic influence of non-adiabatic effects on the dynamics. Starting with the numerically exact result within ML-X, we observe significant deviations for both the bath, see Fig.~\ref{fig:one_body_density_g50_xs15}(a\(_1\)), and the impurity, see Fig.~\ref{fig:one_body_density_g50_xs15}(b\(_1\)), species compared to the previously analyzed cases. Unlike the lower interaction strength in Fig.~\ref{fig:one_body_density_g25_xs15} (\(g = 2.5 \hbar \omega_B \ell_B\)) or a smaller shift (\(x_s = 0.5 \ell_B\)) in Fig.~\ref{fig:one_body_density_g50_x05}, in the case of Fig.~\ref{fig:one_body_density_g50_xs15}(b\(_1\)) the impurity's one-body density is largely unable to penetrate the bath species and it is entirely reflected, performing a small amplitude oscillation near the upper boundary of the bath density (\(x_I \approx 2.5 \ell_B\)). This oscillation occurs locally and with a higher frequency, than the one set by $\omega_I$ \textcolor{black}{as evidenced by the green dotted line}. During the  oscillation dynamics, a small fraction of the impurity density can enter the bath density.  

Consequently, the bath species exhibit a notably different behavior: it maintains its initial shape, characterized by a maximum density at the center, and does not develop pronounced density valleys in the one-body density due to the interaction. However, the bath species perform collective dipole oscillations in response to the impurity, \textcolor{black}{as indicated by the time evolution of the average mean position shown in the plot}. Once a sizable fraction of the impurity density is able to penetrate into the bath density at $t \approx 10 \omega_B^{-1}$, the latter starts exhibiting pronounced minima due to interaction.  
\begin{figure*}
    \centering
    \includegraphics[width=0.8\linewidth]{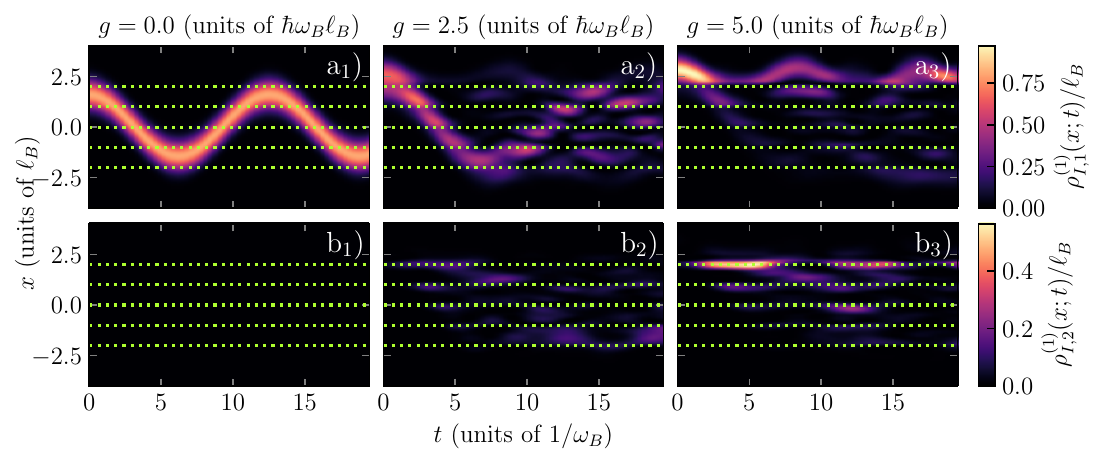}
    \caption{Time dependence of the PEC-resolved one-body density of the impurity, $\rho_{I,n}^{(1)}(x;t)$, corresponding to the $n$-th PEC and for varying interaction strengths (see label). The impurity mass is set to $m_I = 4m_B$ and the trapping frequency to $\omega_I = 0.5\omega_B$, with a larger shift of $x_s = 1.5\ell_B$ taken into account. The green dashed lines indicate the avoided crossings between the ground-state and first-excited PEC, as described in Eq.~\eqref{eqn:potential_energy_surfaces}.
}
    \label{fig:pec_occ_xs_15}
\end{figure*}

The NVABO approximation does not capture this behavior, but rather it exhibits similar dynamics to the previously analyzed cases, see Fig.~\ref{fig:one_body_density_g50_xs15}(a\(_2\)) and \ref{fig:one_body_density_g50_xs15}(b\(_2\)). Qualitatively the behavior is almost identical to the $g = 2.5~\hbar \omega_B \ell_B$ case of Fig.~\ref{fig:one_body_density_g25_xs15}(a\(_2\)) and \ref{fig:one_body_density_g25_xs15}(b\(_2\)). The main differences are a longer oscillation period and the emergence of more pronounced peak density fluctuations on the impurity species, see Fig.~\ref{fig:one_body_density_g50_xs15}(b$_2$).
On the other hand, the VABO approximation, is able to capture some of the main features of the exact impurity dynamics, compare Fig.~\ref{fig:one_body_density_g50_xs15}(a\(_3\)) and \ref{fig:one_body_density_g50_xs15}(b\(_3\)) to Fig.~\ref{fig:one_body_density_g50_xs15}(a\(_1\)) and \ref{fig:one_body_density_g50_xs15}(b\(_1\)). This leads to the conclusion that the trapping of the impurity density outside its bath, that can be seen in Fig.~\ref{fig:one_body_density_g50_xs15}(b\(_3\)), is substantially influenced by the inclusion of the Born-Huang correction, which accounts for changes in the bath's kinetic energy. 
Only a small fraction of the impurity density passes and transmits through the bath species, where it shows tunneling-like dynamics. Nevertheless, the majority of the impurity density remains outside the bath and does not undergo well-pronounced oscillations. As a result, in the bath one-body density (shown in Fig.~\ref{fig:one_body_density_g50_xs15}(a$_3$)), we observe only a slight density valley during the first half of the initial faint-dipole oscillation. This valley subsequently vanishes, leaving behind only minor perturbations (small wiggles) in the bath density. A collective dipole oscillation, as observed in the numerically exact results, see Fig.~\ref{fig:one_body_density_g50_xs15}(a$_1$) is absent for $\rho^{(1)}_B(x;t)$.  

The above already indicates that in this regime the dynamics of the system tends towards being completely diabatic since the impurity is able to excite a collective mode of the bath, which of course it is difficult to accurately describe with a small number of PEC, let alone a single one as in the adiabatic approximation. Nevertheless, comparing to VABO we can verify that the origin of the localization of the impurity outside its fermionic environment \textcolor{black}{originates from the bath's kinetic energy contribution, as accounted for by the Born-Huang term.} \textcolor{black}{Fig.~\ref{fig:impurity_dynamics}(b) offers a graphical illustration of this interplay, showing the qualitative behavior of both the Born-Huang term, which is shifted by a constant factor, and the PECs. It provides an intuitive summary of the spatial dynamics of the impurity, as revealed by the various approaches discussed earlier. The alignment between the positions of the avoided crossings and the peaks of the Born-Huang term $V_{\rm ren
}^{11}(x_I)$ indicates the emergence of an effective harmonic oscillator potential, which leads to the self-trapping of the impurity in the outer region, exhibiting quasi-diabatic behavior dictated by the structure of the PECs. As a result, only a small fraction of the impurity density, as previously observed, is able to escape this trapping.}

\subsubsection{Analysis in terms of PEC}

Building on our earlier analysis of smaller shifts, we once again examine the populations of the PECs to deepen our understanding of the one-body density trends shown in Fig.~\ref{fig:one_body_density_g00}, \ref{fig:one_body_density_g25_xs15}, and \ref{fig:one_body_density_g50_xs15}. This allows us to further unravel the role of beyond Born-Oppenheimer effects, with particular emphasis on the PJTE.
To achieve this, we analyze the contributions of individual PECs via Eq.~\eqref{eqn:pes_occ}, using Fig.~\ref{fig:sum_plot_occ_15} to illustrate their behavior and its implications.
In the simple non-interacting case, as shown in Fig.~\ref{fig:sum_plot_occ_15}(a), we observe, as expected, that only the lowest PEC is involved. This corroborates our findings on the one-body density (Fig.~\ref{fig:one_body_density_g00}(b$_2$)), confirming that this behavior is entirely described within the adiabatic approximation.  

The analysis becomes more intriguing when interactions are introduced, as depicted in Fig.~\ref{fig:sum_plot_occ_15}(b). The contributions from PECs beyond the Born-Oppenheimer approximation increase significantly: the population of the ground-state PEC diminishes, while the populations of the first and second excited PECs rise. In stark contrast to the small shift case, see Fig.~\ref{fig:sum_plot_occ_05}(b) we observe that the population transfer to the excited PEC is very pronounced. Indeed, similarly to the case of $x_s = 0.5~\ell_B$ the first excited PEC has a small population at $t = 0$ due to the influence of the synthetic conical intersections, however, during the time-evolution the population of the ground PEC gets irreversibly (for the time scales considered here) depleted in favor of the excited states. In particular, the occupation of the $j =1$ PEC reaches a minimum of $n_1 \approx 0.6$ at $t \approx 5.5 \omega_B^{-1}$ and then it performs small-amplitude oscillations obeying $n_1 < 0.7$. The interpretation of this behavior is that the momentum transfer caused by the first crossing of the impurity through its fermionic environment results to the generation of a collective excitation of the fermionic environment, see Fig.~\ref{fig:one_body_density_g25_xs15}(a$_1$), which distributes the impurity energy to multiple degrees of freedom of the fermionic bath and thus the initial state of the impurity is no longer able to revive within such a small time frame. 
\begin{figure*}
    \centering
    \includegraphics[width=0.9\linewidth]{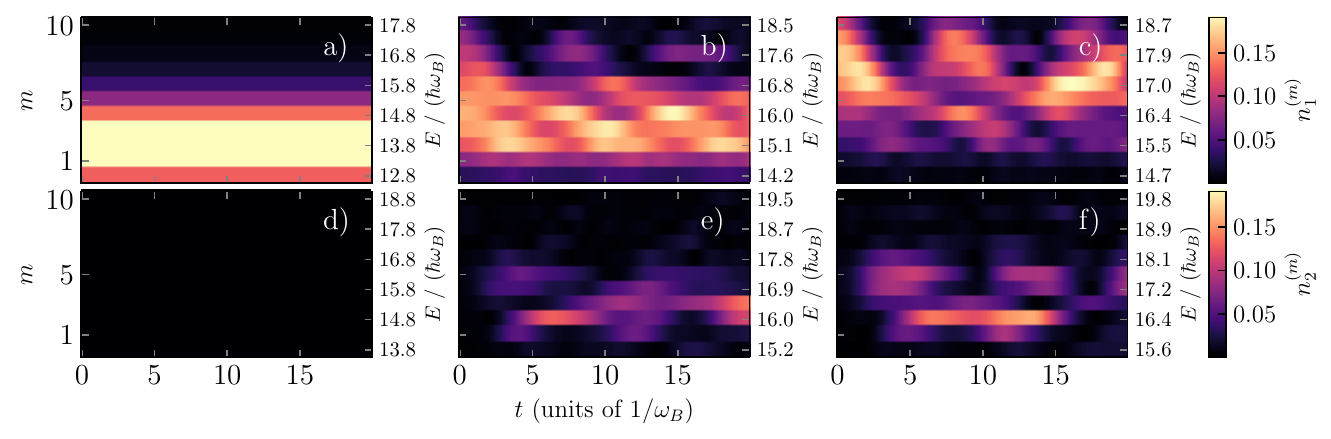}
    \caption{Time-evolution of the population of the eigenstates of (a), (b), (c) the ground-state PEC $j=1$ and (d), (e), (f) the first excited $j=2$ for $m=0,\cdots,10$, see Eq.~\eqref{eqn:states_m_pes_j_ev}. In all cases a shift $x_s=1.5$ is considered. The interaction strength is (a) and (d) $g=0.0$,  (b) and (e) $g=2.5$, and (c) and (f) $g=5.0$. The right vertical axes show the eigenenergy of the corresponding state for the effective single-body Hamiltonian given in Eq. \eqref{eqn:one_body_ham}.}
    \label{fig:2D_plot_occ_15}
\end{figure*}

In the case of strong interactions (Fig.~\ref{fig:sum_plot_occ_15}(c)), we observe a quite different behavior of the ground-state PEC population than the $g = 2.5~\hbar \omega_B\ell_B$ case, as the population of all $j = 1, 2, 3$ states exhibit a pronounced oscillatory behavior that can be associated to the dipole mode of the bath observed in Fig.~\ref{fig:one_body_density_g50_xs15}(a$_1$) as the minima of $n_1$ correspond to the maximum negative displacement of the bath species and the maxima of $n_1$ to the corresponding positive displacements. Of course, the amplitude of the PEC population transfer is non-constant in time, which is justified by the fact that the dipole motion of the bath is accompanied by other type of excitations of this species.

To proceed further, we examine the impurity density contributions of the different PECs in Fig.~\ref{fig:pec_occ_xs_15}. Of course, for $g = 0$ our results are straightforward, since only the $j = 1$ PEC is involved and thus $\rho^{(1)}_{I,1}(x;t) = \rho^{(1)}_{I}(x;t)$, see Fig.~\ref{fig:pec_occ_xs_15}(a$_1$) and  Fig.~\ref{fig:pec_occ_xs_15}(b$_1$). In the weakly interacting case, we observe a qualitatively similar behavior to the case of smaller displacements. Indeed, $\rho^{(1)}_{I,1}(x;t)$ exhibits a structure similar to the impurity density within VABO, as seen in Fig.~\ref{fig:pec_occ_xs_15}(a$_2$) to ~\ref{fig:one_body_density_g25_xs15}(b$_3$), however, the amplitude of $\rho^{(1)}_{I,1}(x;t)$ is significantly reduced due to the property $n_1(t) = \int {\rm d}x~\rho^{(1)}_{I,1}(x;t)$ of the PEC densities and the drastic reduction of $n_1(t)$ during the dynamics, see Fig.~\ref{fig:sum_plot_occ_15}(b). Turning to the case of the excited PEC we observe that the case of $x_s = 1.5 \ell_B$ shows a similar qualitative behavior as the case of $x_s = 0.5 \ell_B$, compare Fig.~\ref{fig:pec_occ_xs_15}(b$_2$) to Fig.~\ref{fig:pec_occ_xs_05}(b$_2$). In particular, a density component of the second PEC emerges near each avoided crossing as the impurity density of the ground PEC goes through this crossing, which indicates diabatic transport. Similarly also to the case of $x_s = 0.5 \ell_B$ this impurity density fraction exhibits tunneling dynamics from one crossing to another, an effect that is more prominently visible for $t \approx 6.5 \omega_B^{-1}$ and $\ell_B \leq x \leq 2 l_B$. The above results suggest that despite the fact that in this case the dynamics of the system deviates substantially from the adiabatic case, via the generation of collective excitations observable in the bath density, the underlying cause is the diabatic transport of the impurity density. This shows that the non-adiabatic behavior of the system observed for both $x_s = 0.5\ell_B$ and $x_s = 1.5\ell_B$ in the $g  = 2.5 \hbar \omega_B \ell_B$ case essentially defines a crossover region from fully adiabatic behavior for very low interactions and the collective excitation behavior for strong interaction, which we examine below.

For strong interactions $g = 5.0 \hbar \omega_B \ell_B$ the behavior of the system changes as Fig.~\ref{fig:pec_occ_xs_15}(a$_3$) and Fig.~\ref{fig:pec_occ_xs_15}(b$_3$) show a very different behavior of the system. Notice that shortly after the dynamics are initiated at $t = 2 \omega_B^{-1}$, the impurity impinges on the $x \approx 2\ell_B$ avoided crossing and very pronounced diabatic transport to the $j = 2$ PEC occurs, transferring the majority of the impurity density to the excited PEC. Then, the density is largely transferred diabatically back to the first excited state at $t \approx 5 \omega_B^{-1}$. This dynamics persists in the examined time-scale. Notice that the remaining adiabatic transport, when combined to the tunneling dynamics in the excited PEC are the reason for the density that penetrates the bath, visible in Fig.~\ref{fig:one_body_density_g50_xs15}(b$_1$).
\begin{figure*}
    \centering
    \includegraphics[width=0.9\linewidth]{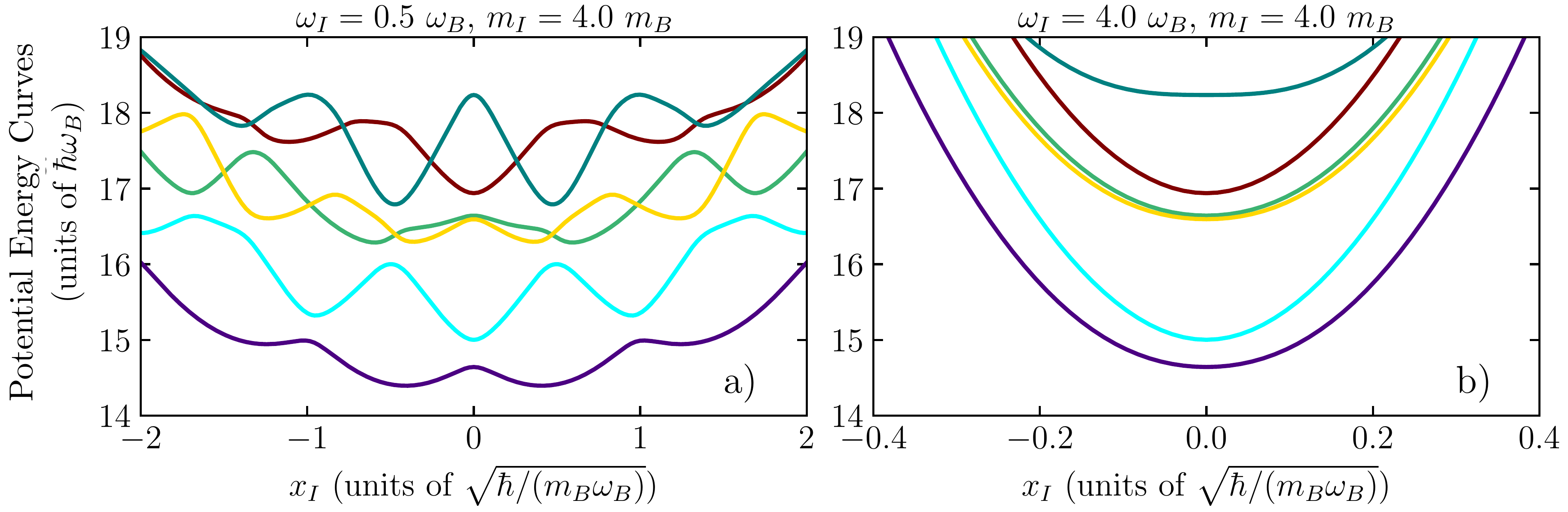}
    \caption{The lowest lying PECs \eqref{eqn:potential_energy_surfaces} from the MCBO approach for an interaction strength $g=5.0~\hbar\omega_B\ell_B$ for a heavy impurity mass $m_I=4m_B$ and the trapping parameters $\omega_I=0.5\omega_B$ in (a) and $\omega_I=4\omega_B$ (b) (see label) ~\cite{Becker2024}.}
\label{fig:potential_energy_surfaces_comparison}
\end{figure*}

The findings of $g = 5 \hbar \omega_B \ell_B$ and $x_s = 1.5 \ell_B$ demonstrate a significant qualitative divergence in behavior than the previous cases. Indeed, the dynamics of the system can be understood more easily by considering the infinitely interacting case of an impenetrable bath. Elementary physics leads to the conclusion that in this case the motion of the impurity after the quench will result to an elastic collision with the bath, where the impurity transfers its momentum to its fermionic environment pushing it along the direction it traveled while the impurity recoils backward, leading to a periodic bath impurity collision every period of the collective oscillation, see also Fig.~\ref{fig:one_body_density_g50_xs15}(a$_1$) and  Fig.~\ref{fig:one_body_density_g50_xs15}(b$_1$). Notice that similar phenomenology has already been observed in Bose-Fermi mixtures~\cite{CaoBolsinger2017}. Although the dynamics can be understood within the language of PEC, the analysis becomes tedious due to the large superposition of PEC that gets involved in the dynamics, owing to the momentum deposition in the bath component. Also the transport through the conical intersections is simple due to the almost perfect diabatic transport in the $g \to \infty$ limit. Therefore, in summary our analysis indicates that the large interaction energy suppresses the interesting non-adiabatic effects observed in the other cases since diabatic transport dominates.

To finish this section we briefly discuss the occupation of different vibrational states of the two lowest PEC. 
In the non-interacting case, the shift in potential energy leads to the occupation of higher-lying states of the ground-state PEC at the center of the trapping potential (Fig.~\ref{fig:2D_plot_occ_15}(a)). With the increased displacement, $x_s = 1.5$ we observe that the first excited state of the ground-state PEC initially becomes the most populated as the Poisson distribution of the eigenstate population within the initial coherent state becomes more apparent. Due to the absence of interactions, the bath and impurity remain separable, resulting in no temporal evolution of the state populations.
Upon introducing weak interactions (\( g=2.5\hbar\omega_B\ell_B \)), as depicted in Fig.~\ref{fig:2D_plot_occ_15}(b) and (d), we initially observe a behavior in the ground-state PEC similar to the non-interacting case but shifted to higher lying eigenstates, owing to the extra-shift of the initial impurity configuration due to the interaction. However, once the system encounters an avoided crossing—evidenced by the still-discernible dipole oscillation in the impurity’s one-body density (Fig.~\ref{fig:one_body_density_g25_xs15}(b$_1$))—the second to fifth eigenstates of the first excited PEC become significantly populated. Notably, in contrast to the smaller displacement case, these states (including the third state) remain occupied over time. The effective potential of the first excited PEC, when combined with the structure of the corresponding eigenfunctions, reveals that the triple-well structure facilitates trapping, whereas the weakly pronounced double-well structure in the ground-state PEC does not. 
Moving to the strongly interacting regime (\( g=5.0\hbar\omega_B\ell_B \)), we observe a similar pattern. In particular, during time-instances where the impurity density penetrates the trap center despite the repulsive interaction, the first excited state in the excited PEC dominates, as shown in Fig.~\ref{fig:2D_plot_occ_15}(c). In contrast, the lowest state in the ground-state PEC remains effectively unoccupied. This behavior correlates with the oscillation of the impurity’s one-body density, which extends beyond \( x>2.0\ell_B \), with only a small fraction reaching the trap center. Examining the ground-state PEC's zeroth eigenfunction (Fig.~\ref{fig:lowest_PES_eigenstates}(b$_3$)), we note its localization at the trap center, yet the strong repulsion prevents significant impurity presence in this region.
Finally, a key point of Fig.~\ref{fig:lowest_PES_eigenstates}(a$_2$), \ref{fig:lowest_PES_eigenstates}(b$_2$), \ref{fig:lowest_PES_eigenstates}(a$_3$), and \ref{fig:lowest_PES_eigenstates}(b$_3$) is that in the interacting case the occupation of states in the excited PEC lie in the same energy range as the lowest lying PEC.

\section{Stronger trapping confinement of the impurity}
\label{sec:Stronger trapping confinement of the impurity}

\begin{figure*}
    \centering
    \includegraphics[width=1.0\linewidth]{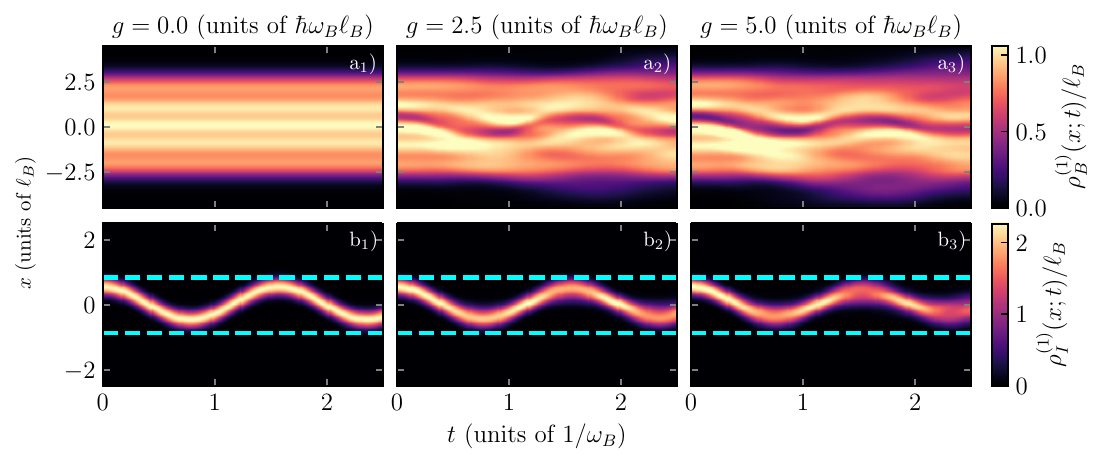}
    \caption{Spatiotemporal dependence of the one-body density of the bath $\rho_B^{(1)}(x;t)$ and impurity $\rho_I^{(1)}(x;t)$ species for varying interaction strengths (see label). The impurity mass $m_I=4m_B$ and stronger trapping frequency $\omega_I=4.0\omega_B$ as well as the shift $x_s=0.5\ell_B$ are considered. For the impurity species, we include cyan dashed lines at $x = \pm(x_s+\sqrt{2}\ell_I)$ as a visual reference to the oscillation amplitude in the non-interacting case.}
    \label{fig:obd_xs05_m4_w4}
\end{figure*}
As part of our ground-state analysis, we explored the dynamics of a tightly confined impurity species in depth \cite{Becker2024}. Building on this foundation, we now examine the dynamic behavior for this scenario $\omega_I=4\omega_B$, focusing on the case where the avoided crossing at $x = 0$ occurring between the lowest and first excited PECs becomes not visible because of the large contribution of the impurity confining potential, compare Fig.~\ref{fig:potential_energy_surfaces_comparison}(b) to~\ref{fig:potential_energy_surfaces_comparison}(a).

To this end, we consider the case of a small displacement ($x_s=0.5\ell_B$) to specifically investigate the behavior near the trap center in Fig.~\ref{fig:obd_xs05_m4_w4}. In the non-interacting case ($g=0$), as expected, no changes are observed in the one-body density of the bath, as depicted in Fig.~\ref{fig:obd_xs05_m4_w4}(a$_1$). For the impurity species in Fig.~\ref{fig:obd_xs05_m4_w4}(b$_1$), aside from a reduced confinement length of the impurity one-body density when compared to the $\omega_I = 0.5 \omega_B$ case examined previously, we observe dipole oscillations occurring on a significantly faster timescale - anticipated given the higher trap frequency (compare to Fig.~\ref{fig:one_body_density_g00}).

When interactions are introduced, the behavior of the bath’s one-body density changes substantially in Fig.~\ref{fig:obd_xs05_m4_w4}(a$_2$): For $g=2.5\hbar\omega_B\ell_B$ a pronounced density dip emerges due to the repulsive interaction, displacing the bath’s density maximum, which was previously centered in the trap in the non-interacting case. This displacement, driven by the impurity, can be interpreted as excitations within the bath. For the impurity species (b$_2$), only minor adjustments are observed. The dipole oscillation remains clearly visible, although interactions and the bath's initial density structure cause a broadening of the density at the turning points and a sharpening near the trap center. Compared to our previous investigations, it is evident that the expected splitting of the impurity density—strongly associated with the PJTE does not occur in this case.

This trend continues in the strongly interacting regime ($g=5.0\hbar\omega_B\ell_B$). Here, the bath’s one-body density Fig.~\ref{fig:obd_xs05_m4_w4}(a$_3$) exhibits a more pronounced central minimum, with significantly stronger displacement of the bath species by the impurity. The bath excitations due to interaction intensify, leading to sharper maxima adjacent to the density minima. At the outer edges of the bath, subtle oscillations emerge, induced by the strong repulsion from the impurity. For the impurity’s one-body density in Fig.~\ref{fig:obd_xs05_m4_w4}(b$_3$), the previously observed effects—density concentration at the center and broadening at turning points—become more pronounced.

The above indicates that the case of stronger confinement strongly affects the behavior of the impurity. Since the dynamics becomes much faster and the impurity caries more density, we observe a largely diabatic behavior of the impurity and the excitation of collective modes as in Fig.~\ref{fig:one_body_density_g50_xs15}(a$_1$) and Fig.~\ref{fig:one_body_density_g50_xs15}(b$_1$). Therefore, the strongly confined impurity case is not particularly attractive for studying beyond Born-Oppenheimer phenomena.

\section{Summary and Outlook}
\label{sec:Summary and Outlook}

We have investigated the real-time dynamics of a mass-imbalanced, fermionic few-body system, highlighting the fundamental role of non-adiabatic effects induced by the PJTE. By employing numerically exact methods, specifically, the Multi-Layer Multi-Configuration Time-Dependent Hartree method and the multi-channel Born-Oppenehimer approach, we have provided a detailed analysis of how conical intersections and avoided crossings influence the system’s evolution.

Our findings reveal that non-adiabatic coupling becomes increasingly significant with stronger impurity-bath interactions in mass-imbalanced systems. Notably, we demonstrated that the conventional Born-Oppenheimer picture breaks down not only in extreme regimes but already at moderate interaction strengths. This breakdown manifests in pronounced population transfer between different potential energy surfaces, driven by pseudo Jahn-Teller effect-induced conical intersections.

A key result is the dependence of non-adiabatic behavior on the initial displacement and quench of the impurity trapping potential. For small displacements, the system is predominantly influenced by a single avoided crossing. However, for larger displacements, multiple avoided crossings come into play, leading to enhanced energy redistribution, self-trapping effects, and a modified impurity transport mechanism. Our analysis shows that impurity dipole oscillations increasingly dephase and become governed by a complex interplay of potential energy surfaces and conical intersections. Nevertheless, when the energy imparted to the bath becomes too large either because of an increased displacement of the impurity trapping frequency, the dynamics of the impurity becomes strongly diabatic involving the collective excitation of the bath. Therefore, suppressing the influence of beyond Born-Oppenheimer effects associated to the crossover between adiabatic and diabatic behavior, as the dynamics becomes amenable to simplified diabatic transport models.

These results provide significant insights into non-adiabatic processes in ultracold quantum gases and establish a theoretical foundation for experimental investigations in controlled atomic settings. The ability to tune interactions via Fano-Feshbach resonances or confinement-induced resonances opens new possibilities for experimentally realizing quantum-controlled impurity dynamics. \textcolor{black}{In particular, \textit{in situ} imaging provides access to both the impurity and bath position, allowing for a direct comparison to the spatial resolved densities in this work.}

Based on our results, several pathways for future research emerge. 
An intriguing direction involves incorporating spin-orbit coupling in the impurity-bath system. Since the Jahn-Teller effect frequently occurs in heavy-element compounds, the presence of spin-orbit interactions in ultracold gases could provide a direct link between our system and molecular physics \cite{Bersuker2021, Bersuker2023}. Moreover, ultracold atom experiments enable spin-resolved measurements, offering a unique opportunity to observe symmetry-breaking processes induced by non-adiabatic dynamics. Encoding the PJTE in the spin-state evolution of the system could provide a powerful experimental tool for probing the interplay between impurity motion and internal degrees of freedom.

Beyond one-dimensional confinement, extending our study to isotropically trapped two- and three-dimensional systems could reveal novel non-adiabatic effects. In such setups, impurity displacement breaks the rotational symmetry of the trap, potentially leading to conical intersections that drive complex quantum dynamics \cite{BuschEnglert1998}. The competition between rotational barriers and interaction energy may lead to novel impurity-induced phase transitions or correlated many-body effects.

Finally, our findings naturally connect to broader concepts in many-body physics. In particular, a fascinating question is the potential link between the non-adiabaticity observed here and the well-known Anderson orthogonality catastrophe \cite{Anderson1967, Anderson1967_2}. 
\textcolor{black}{Traditionally, the Anderson orthogonality catastrophe arises when a localized and strongly interacting impurity modifies the many-body wavefunction of its surrounding bath, leading to a vanishing overlap between the interacting and non-interacting ground states in the thermodynamic limit, that is \(Z = \sqrt{|\langle \psi_{g=0} | \psi_{g \neq 0} \rangle|^2} \propto N^{-\alpha}\) with \(\alpha > 0\). In our framework, this orthogonality is inherently present from the outset: the multi-channel Born-Oppenheimer ansatz explicitly accounts for the dependence of each bath fermion's state on the instantaneous position of the impurity. These couplings induce collective excitations and lead to spontaneous symmetry breaking and impurity self-trapping.
This constitutes a qualitatively distinct regime of impurity physics, where the breakdown of adiabaticity gives rise to emergent, non-perturbative dynamics beyond the conventional polaron picture, in which the impurity remains delocalized and is dressed by virtual particle hole excitations~\cite{Chevy2006}. In contrast, the pseudo-Jahn-Teller regime features strong, non-adiabatic coupling that results in localized impurity behavior and symmetry-broken configurations. Our results thus provide a foundation for future studies on quasiparticle breakdown and many-body entanglement arising from non-adiabatic impurity motion, offering new insights into quantum impurity problems in ultra-cold atomic systems and their collective behavior.
}

By exploring these directions, future studies can deepen our understanding of non-adiabatic quantum dynamics and establish ultracold atomic systems as a versatile platform for probing impurity physics beyond the Born-Oppenheimer approximation.

\section*{Acknowledgements}

This work has been funded by the Cluster of Excellence “Advanced Imaging of Matter” of the Deutsche
Forschungsgemeinschaft (DFG) - EXC 2056 - project ID 390715994. G.K.M. has received funding by the Austrian Science Fund (FWF) [DOI: 10.55776/F1004].

\appendix
\section{The ML-X method}
\label{app:MLX}

At the beginning in Sect.~\ref{sec:Methodology and Computation Approach}, we provide a concise overview of the employed numerical approaches. The purpose of this section is to delve into the ML-X method.
The ML-X ansatz expresses the total many-body wavefunction, $|\Psi(t)\rangle$, as a linear combination of $j=1,2,\dots,D$ distinct orthonormal functions for each involved species
\begin{equation}
    |\Psi(t)\rangle = \sum_{j_B, j_I = 1}^D A_{j_B, j_I}(t) |\Psi_{j_B}^B(t)\rangle |\Psi_{j_I}^I(t)\rangle,
\end{equation}
where $|\Psi_j^\sigma(t)\rangle$ ($\sigma = B, I$) are species wavefunctions, and $A_{j_B, j_I}(t)$ are time-dependent expansion coefficients. This decomposition is formally equivalent to a Schmidt decomposition of rank $D$
\begin{equation}
    |\Psi(t)\rangle = \sum_{k=1}^D \sqrt{\lambda_k(t)} |\tilde{\Psi}_k^B(t)\rangle |\tilde{\Psi}_k^I(t)\rangle,
\end{equation}
considering the $\lambda_k(t)$ (Schmidt weights) as eigenvalues of the reduced density matrix $\rho_\sigma^{(N_\sigma)}(t)$, and the eigenstates $|\tilde{\Psi}_k^\sigma(t)\rangle$ as Schmidt modes.
Hereby, the reduced density matrix for species $\sigma$ is given by
\begin{equation}
\begin{split}
\rho_{\sigma}^{(N_\sigma)}&(x_1, \dots, x_{N_\sigma}, x'_1, \dots, x'_{N_\sigma}, t)=
\int \prod_{j = 1}^{N_{\bar \sigma}} \mathrm{d} x_j^{\bar \sigma}~ \\
&\times \Psi^*(x_1^{\sigma}=x'_1, \dots, x_{N_\sigma}^{\sigma}=x'_{N_\sigma}, x^{\bar \sigma}_1, \dots, x^{\bar \sigma}_{N_{\bar \sigma}}, t) \\
&\times \Psi(x_1^{\sigma}=x_1, \dots, x_{N_\sigma}^{\sigma}=x_{N_\sigma}, x^{\bar \sigma}_1, \dots, x^{\bar \sigma}_{N_{\bar \sigma}}, t),
\end{split}
\end{equation}
with $\bar{\sigma} \neq \sigma$. Here $N_{\sigma}$ and  $N_{\bar{\sigma}}$ denotes the number of atoms belonging to the respective species $\sigma$, $\bar{\sigma}$. Within the ML-X framework, the density matrix operator can be expanded as  $\rho_{\sigma}^{(N_\sigma)}(x_1, \dots, x_{N_\sigma}, x'_1, \dots, x'_{N_\sigma}, t)=\langle x_1, \dots, x_{N_\sigma}  | \hat{\rho}^{(N_{\sigma})}_{\sigma}(t)| x'_1, \dots, x'_{N_\sigma} \rangle$ with the density matrix operator 
\begin{equation}
\hat{\rho}_\sigma^{(N_\sigma)}(t) = \sum_{j_\sigma,j'_\sigma=1 \atop j_{\bar{\sigma}}=1}^D \left[\hat{\rho}_\sigma^{(N_\sigma)}(t)\right]_{j_\sigma,j'_\sigma} |\Psi_{j_\sigma}^\sigma(t)\rangle \langle\Psi_{j'_\sigma}^\sigma(t)|
\end{equation}
introducing $A_{j_{\sigma}, j_{\bar{\sigma}}}^*(t) A_{j_{\bar{\sigma}}}, j'_{\sigma}(t)\equiv \left[\hat{\rho}_{\sigma}^{(N_\sigma)}(t)\right]_{j_{\sigma}, j'_{\sigma}}$.
Hence, diagonalizing $\left[\hat{\rho}_\sigma^{(N_\sigma)}(t)\right]_{j_\sigma,j'_\sigma}$ for $j_{\sigma},j_{\sigma}'=1,\cdots,D$ yields $\lambda_k(t)$ and $|\tilde{\Psi}_k^\sigma(t)\rangle$.\\\\
The key feature of the ML-X method is the multilayered structure. It arises from the expansion of each species wavefunction, $|\Psi_j^\sigma(t)\rangle$ in terms of time-dependent number states $|\vec{n}(t)\rangle^\sigma$ leading to
\begin{equation}
    |\Psi_j^\sigma(t)\rangle = \sum_{\vec{n}} B_{j,\vec{n}}^\sigma(t) |\vec{n}(t)\rangle^\sigma,
\end{equation}
where $B_{j,\vec{n}}^\sigma(t)$ corresponds to time-dependent expansion coefficients in the single-species number state basis, $|\vec{n}(t)\rangle^\sigma$. These number states are built in terms of $d^\sigma$ time-dependent variationally optimized Single-Particle Functions (SPFs) given by $\phi_l^\sigma(t)$, $l=1,2,..., d^\sigma$ with $\vec{n}=(n_1,...,n_{d^{\sigma}})$ corresponding to the number of atoms in each SPF. On the lowest layer, the SPFs are expanded in a time-independent DVR basis $\{|k\rangle\}$ and are defined as
\begin{equation}
    |\phi_j^\sigma(t)\rangle = \sum_{k=1}^{\mathcal{M}} C_{jk}^\sigma(t) |k\rangle.
\end{equation}
In this study, we used the points on the DVR grid $\mathcal{M} = 150$ of a harmonic oscillator.\\\\
To compute the ground state at the beginning of our analysis, imaginary-time propagation is performed using $\tau = -it$. This causes the energy of the state to decay proportionally to $e^{-(E(t)-E_0)t}$, converging to the ground state ($E_0$) as $\tau \to \infty$.\\\\
The ansatz’s Hilbert space truncation is characterized by the choice of the orbital configuration space, which is represented by $C = (D; d^B; d^I)$, where we choose the following set for our investigation \cite{Becker2024}:
\begin{itemize}
    \item $d^B = 18$: Bath orbitals to capture the intra-species bath correlations
    \item $d^I = 12$: 
    Are found to be enough for convergence of the impurity species
    \item $D = d^I = 12$: Incorporating all possible Schmidt modes of inter-species entanglement, for given $d^I$. 
\end{itemize}
This choice is based on an extensive convergence analysis for our ground state, which has also demonstrated its validity in the dynamics, as shown by examining the occupation numbers of the respective orbitals.
The equations of motion are derived using the Dirac-Frenkel variational principle \cite{Dirac1930annihilation, frenkel1934wave}
\begin{equation}
    \langle \delta \Psi(t)| i\hbar \frac{\partial}{\partial t} - H |\Psi(t)\rangle = 0.
\end{equation}
This leads to $D^2$ linear differential equations for $A_{j_B,j_I}(t)$, coupled to nonlinear integro-differential equations for $B_{j,\vec{n}}^\sigma(t)$ and $C_{j,k}^\sigma(t)$.
\section{Derivation of potential energy surface diagnostics}
\label{app:Derivations in Second Quantization Formalism}

The multi-channel Born-Oppenheimer ansatz of Eq.~\eqref{eqn:multi-channel_BornOppenheimer} when casted in the second quantization formalism reads
\begin{equation}
|\Psi (t)\rangle = \int \mathrm{d}x_I \, \Psi_{j,I}(x_I; t) \hat{\Psi}_I^\dagger(x_I) |0_I\rangle \otimes |\Psi_{j,B}(x_I)\rangle,
\end{equation}
where \( \Psi_{j,I}(x_I; t) \) is the multi-channel impurity wavefunction and is the only dynamical variable of the above many-body state. Further, \( |\Psi_{j,B}(x_I)\rangle \) describes the state of the bath when the impurity is at a fixed position, which as discussed in Sec.~\ref{mcBO_main_text} is the eigenstate of the bath Hamiltonian $\hat{H}_B + \hat{H}_{BI}$ when $x_I$ is treated as an external parameter. Finally, \( \hat{\Psi}_I^\dagger(x_I) \) creates an impurity particle at \(x_I\), obeying the standard fermionic anti-commutation relation $\{\hat{\Psi}_I(x_1), \hat{\Psi}_I^\dagger(x_2)\} = \delta(x_1 - x_2)$ with the corresponding annihilation operator,  \( \hat{\Psi}_I(x_I) \).

In this formalism, the one-body density matrix of the impurity is defined as
\begin{equation}
\rho_I^{(1)}(x_1,x_2; t) = \langle \Psi(t)| \hat{\Psi}_I^\dagger(x_1) \hat{\Psi}_I(x_2) |\Psi(t)\rangle,
\end{equation}
which after some straightforward fermionic anti-commutation algebra reads
\begin{equation}
\begin{split}
    \rho_I^{(1)}(x_1,x_2;t) = \sum_{j,k=1}^M& \Psi_{j,I}^*(x_1;t)\Psi_{k,I}(x_2;t) \\ &\times \langle \Psi_{j,B}(x_1) | \Psi_{k,B}(x_2) \rangle.
\end{split}
\end{equation}
This shows that the one-body density matrix contains contributions from overlaps between bath states \( \langle \Psi_{j,B}(x_1) | \Psi_{k,B}(x_2) \rangle \), weighted by the corresponding impurity amplitudes. Notice that the one-body density $\rho^{(1)}_I(x_I;t) = \rho^{(1)}(x_I, x_I;t)$ further simplifies to 
\begin{equation}
\begin{split}
    \rho_I^{(1)}(x_I;t) &= \sum_{j = 1}^M \underbrace{|\Psi_{j,I}(x_I;t)|^2}_{\equiv \rho_{I,j}^{(1)}(x_I;t)},
\end{split}
\label{one_body_density_components}
\end{equation}
where we have defined $\rho_{j, I}^{(1)}(x_I;t)$ as the PEC specific density contribution. The overlap between bath states is simplified because $\langle \Psi_{j,B}(x_I) | \Psi_{k,B}(x_I) \rangle = \delta_{j,k}$. 

To derive the probability of occupying a certain PEC curve $j$, we introduce the projector 
\begin{equation}
\hat{\mathcal{P}}_j = \int \mathrm{d}x_I \, |\Psi_j(x_I)\rangle \langle \Psi_j(x_I)|,
\end{equation}
where $|\Psi_j(x_I)\rangle$ are defined as
\begin{equation}
|\Psi_j(x_I)\rangle = \hat{\Psi}_I^\dagger(x_I)|0_I\rangle \otimes |\Psi_{j,B}(x_I)\rangle.
\end{equation}
The physical significance of this projector is that it sums over all localized states of the impurity, $|\Psi_j(x_I)\rangle$, upon which the PEC is unambiguously defined. It can be easily verified that it possesses all the properties of a projector, i.e. $\hat{\mathcal{P}}_{j}\hat{\mathcal{P}}_{k} = \delta_{j,k} \hat{\mathcal{P}}_{j}$ and $\sum_{j = 1}^M \hat{\mathcal{P}}_{j} = \hat{I}_{M}$, with $\hat{I}_M$ the identity operator within the truncated basis of the $M$ lowest in energy PEC. These properties can be straightforwardly proven by noting that $\langle\Psi_j(x_1)|\Psi_k(x_2)\rangle = \delta_{j,k} \delta(x_1 - x_2)$, stemming from fermionic anti-commutation relations and $\langle \Psi_{j,B}(x_I) | \Psi_{k,B}(x_I) \rangle = \delta_{j,k}$ (see discussion below Eq.~\eqref{one_body_density_components}).
In the next step, we calculate the expectation value of the projector
\begin{equation}
\begin{split}
    \langle \Psi (t) | \hat{\mathcal{P}}_j | \Psi (t) \rangle &= \int \mathrm{d}x_I \, |\Psi_{j,I}(x_I;t)|^2 \\&= \int \mathrm{d}x_I \, \rho^{(1)}_{I,j}(x_I;t),
\end{split}
\label{projector_in_PEC}
\end{equation}
where we have again used $\langle \Psi_{j,B}(x_I) | \Psi_{k,B}(x_I) \rangle = \delta_{j,k}$.
The expression above shows that the probability of occupying a particular state on the PEC depends solely on the norm of the impurity wavefunction in the $j$-th channel integrated over all space. In addition, Eq.~\eqref{projector_in_PEC} justifies {\it a posteriori} the definition of $\rho^{(1)}_{j, I}(x_I)$ as the contribution of the $j$-th PEC to the one-body density in Eq.~\eqref{one_body_density_components}, since the integral of this density function gives the occupation of the corresponding PEC.

Moreover, by focusing on the different occupied states of the $j$-th PEC, we can define the projector to a specific impurity state \(| \phi_{m, j} \rangle\) of it
\begin{equation}
\hat{\mathcal{P}}_{m,j} = |\Psi_{m,j}\rangle \langle \Psi_{m,j}|.
\end{equation}
The corresponding state in this case is defined as 
\begin{equation}
|\Psi_{m,j}\rangle = \int \mathrm{d}x_I \, \phi_{m,j}(x_I) \hat{\Psi}_I^\dagger(x_I) |0_I\rangle \otimes |\Psi_{j,B}(x_I)\rangle,
\end{equation}
where $\phi_{m,j}(x_I)$ is the wavefunction of the state we want to obtain the occupation of. The operators $\hat{\mathcal{P}}_{m,j}$ obeys all the proper projector properties, provided that the $| \phi_{m, j} \rangle$ states for varying $m$ form an orthonormal basis, the particular property $\hat{\mathcal{P}}_{m,j}^2 = \hat{\mathcal{P}}_{m,j}$ only requires that the state is normalized $\int {\rm d}x_I~|\phi_{m,j}(x_I)|^2 = 1$.
By incorporating the above assumptions the expectation value of the projector $\hat{\mathcal{P}}_{m,j}$ reads
\begin{equation}
\langle \Psi (t) | \hat{\mathcal{P}}_{m,j} | \Psi (t) \rangle = \left| \int \mathrm{d}x_I \, \Psi_{I,j}^*(x_I;t) \phi_{m,j}(x_I) \right|^2,
\end{equation}
notice that $\langle \Psi_{j,B}(x_I) | \Psi_{k,B}(x_I) \rangle = \delta_{j,k}$ was used for this derivation (see discussion below Eq.~\eqref{one_body_density_components}).
Therefore the occupation of a given state \( \phi_{m,j}(x_I) \) of the $j$-th PEC in the many-body state, is just the squared overlap of $j$-th PEC component of the MCBO wavefunction $\Psi_{j,I}^*(x_I;t)$ with the wavefunction of the target state $\phi_{m,j}(x_I)$. 

\bibliography{cleaned_bibliography}
\end{document}